\pdfoutput=1

\documentclass[11pt]{article}

\usepackage[preprint]{acl}

\usepackage[T1]{fontenc}

\usepackage{amsmath}
\usepackage{amsthm}
\usepackage{bbm}
\usepackage{wrapfig,lipsum}
\usepackage{xspace}
\usepackage{booktabs,cellspace}  
\usepackage{microtype}

\usepackage{inconsolata}

\usepackage{graphicx}

\usepackage{colortbl}  
\usepackage{xcolor}  
\usepackage{amsmath}
\usepackage[utf8]{inputenc}
\usepackage[most]{tcolorbox}
\usepackage{pgfplots}
\usepackage{enumitem}
\usepackage[table]{xcolor}
\usepackage{latexsym}

\usepackage{pifont}
\usepackage{booktabs}
\usepackage{amssymb}
\usepackage{times}
\usepackage{adjustbox}
\usepackage{multirow}
\usepackage{caption}
\usepackage{subcaption}

\usepackage{float}
\definecolor{nBlue}{RGB}{0,165,249}
\definecolor{nGreen}{rgb}{0, 0.5, 0.2}
\definecolor{nRed}{rgb}{0.8, 0.1, 0.2}
\definecolor{mGreen}{rgb}{0.3, 0.65, 0.4}
\usepackage[ruled, vlined, noend, linesnumbered]{algorithm2e}

\title{Towards Provably Secure Generative AI: Reliable Consensus Sampling}

\author{
\textbf{Yu Cui}\textsuperscript{1} \quad
\textbf{Hang Fu}\textsuperscript{1} \quad
\textbf{Sicheng Pan}\textsuperscript{1} \quad
\textbf{Zhuoyu Sun}\textsuperscript{1}  \\
\textbf{Yifei Liu}\textsuperscript{1} \quad
\textbf{Yuhong Nie}\textsuperscript{1} \quad
\textbf{Bo Ran}\textsuperscript{1} \quad
\textbf{Baohan Huang}\textsuperscript{1} \quad 
\textbf{Xufeng Zhang}\textsuperscript{1} \\
\textbf{Haibin Zhang}\textsuperscript{2} \quad
\textbf{Cong Zuo}\textsuperscript{1} \quad
\textbf{Licheng Wang}\textsuperscript{1} 
\\ 
\textsuperscript{1}Beijing Institute of Technology \\
\textsuperscript{2}Yangtze Delta Region Institute of Tsinghua University, Zhejiang \\
\texttt{cuiyu@bit.edu.cn, bchainzhang@aliyun.com}
}

\pgfplotsset{compat=1.18}
\begin{document}
\maketitle
\begin{abstract}
Existing research on generative AI security is primarily driven by mutually reinforcing attack and defense methodologies grounded in empirical experience. This dynamic frequently gives rise to previously unknown attacks that can circumvent current detection and prevention. This necessitates the continual updating of security mechanisms. Constructing generative AI with provable security and theoretically controllable risk is therefore necessary. Consensus Sampling (CS) is a promising algorithm toward provably secure AI. It controls risk by leveraging overlap in model output probabilities. However, we find that CS relies on frequent abstention to avoid unsafe outputs, which reduces utility. Moreover, CS becomes highly vulnerable when unsafe models are maliciously manipulated. To address these issues, we propose a new primitive called Reliable Consensus Sampling (RCS), that traces acceptance probability to tolerate extreme adversarial behaviors, improving robustness. RCS also eliminates the need for abstention entirely. We further develop a feedback algorithm to continuously and dynamically enhance the safety of RCS. We provide theoretical guarantees that RCS maintains a controllable risk threshold. Extensive experiments show that RCS significantly improves robustness and utility while maintaining latency comparable to CS. We hope this work contributes to the development of provably secure generative AI.
\end{abstract}

\section{Introduction}

\begin{figure}[t]
    \centering
    \includegraphics[width=0.97\linewidth]{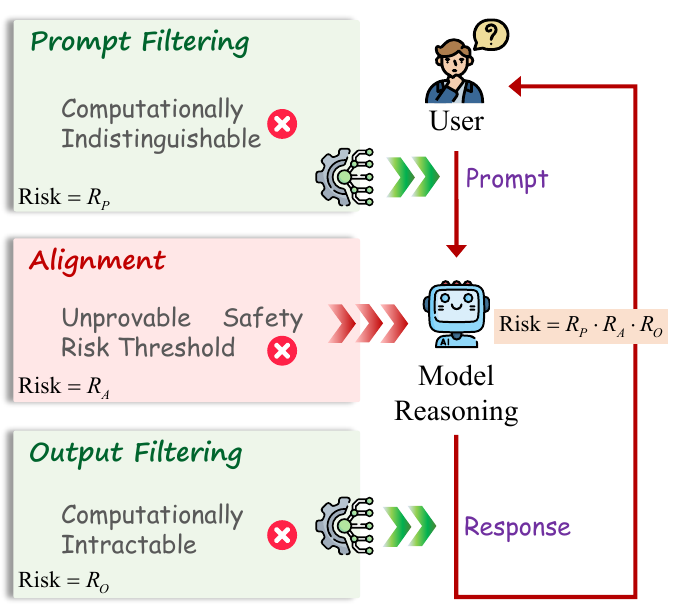}
    \caption{Overview of security risks in generative model reasoning. These risks arise from the aggregation of risks across three core stages. From a theoretical perspective, such risks are unavoidable.}
    \label{fig:example}   
    \vspace{-15pt}
\end{figure}

With the widespread deployment of generative AI, especially large language models (LLMs), security issues continue to emerge \citep{ji2023beaver, Liu2024injection, Zhan2025CAIS, wang2025uniquesec}. Current AI safety research largely follows a coevolutionary trajectory between attacks and defenses \citep{zhang2025jbshield}. New attack methods and defenses appear continuously. This dynamic leaves existing defenses unprepared for future and unpredictable threats. As a result, the definition of AI safety requires constant revision and expansion. The root cause is inherent risk in model reasoning. This risk is unavoidable and difficult to control, as shown in \autoref{fig:example}. Across the full reasoning pipeline, three stages permit risk intervention. At the model level, emerging threats demand repeated alignment. However, safety fine-tuning \citep{jain2024makes} introduces additional risks, such as backdoor attacks \citep{xu-etal-2024-instructions, wen2024privacy} and data poisoning \citep{chen2024agentpoison}. For external filtering mechanisms, prior work \citep{ball2025impossibility} based on cryptographic hardness shows that efficient prompt filters do not exist for certain large models. Output filtering is computationally intractable. Moreover, some risks remain undetectable \citep{kalai2025consensus}. Therefore, security risk in model reasoning cannot be eliminated. Empirical defenses offer only temporary protection. A generative AI paradigm with provable and controllable risk is therefore necessary. Current research on provably safe AI remains position oriented \citep{dalrymple2024towards}.

Recent work on consensus sampling (CS) \citep{kalai2025consensus} offers partial theoretical control over output risk. CS considers a model group with $s$ safe models and $f$ unsafe models ($s > f$). It exploits overlap among output distributions across models to set the acceptance probability for a response of unknown safety. The goal is to make the delivered response more likely to originate from safe models. The aggregated risk of the group admits a provable upper bound. Unlike prior approaches, CS does not define a safe response. It follows the principle that responses supported by more models carry lower risk. This line of work is orthogonal to empirical safety optimization.

However, CS relies on frequent abstention to avoid unsafe outputs, reducing utility \citep{paulus2025safety}. We further show that adversarial control over unsafe models sharply weakens CS security. Despite a cryptographic upper bound on risk, CS lacks robustness and practicality for deployment.

To address these issues, we propose a new primitive named Reliable Consensus Sampling (RCS). We present a provable safety framework for model groups that includes safety and liveness properties. The definitions draw on classical reliable distributed consensus theory. They enable formal security analysis for model groups. RCS records acceptance probability in real time after sampling failures. After a bounded number of rejections, it enters a trace phase that guarantees eventual delivery of a response. RCS fully avoids abstention. During the trace phase, it reweights model probability distributions to control their influence on the final decision. This design tolerates more extreme adversarial behavior and improves robustness. In addition, inspired by quantum entanglement \citep{nielsen2010quantum}, we introduce a feedback algorithm that captures correlations among model distributions. The algorithm identifies models that are unsafe for sampling on specific tasks. It improves RCS safety by excluding those models from the group decision.

We prove that RCS achieves a controllable risk threshold. Extensive experiments show that RCS significantly improves security and utility compared with CS. Latency remains comparable to CS. We summarize our contributions as follows:

\begin{itemize}[left=0pt, itemsep=0pt]

\item We define a security property theory for model groups under a Byzantine threat model. The theoretical framework supports provable security analysis.

\item We propose RCS, a trace-based method that eliminates abstention and guarantees eventual delivery. It tracks acceptance probabilities in real time, reweights models to mitigate adversarial influence, and uses a feedback module that exploits cross-model correlations to improve safety.

\item We theoretically prove that RCS admits a tight upper bound on risk. Experiments demonstrate that RCS outperforms CS in robustness and utility while maintaining comparable latency.

\end{itemize}

\section{Preliminary Analysis}

\noindent
\textbf{Notation}.
We follow the notation established in~\citep{kalai2025consensus}. For a prompt $x$, each model $i \in \{1, \dots, n\}$ induces a probability distribution $p_i(y)$ over a output space $\mathcal{Y}$. Let $\mathrm{Distr}(\mathcal{Y})$ denote the set of probability distributions on $\mathcal{Y}$. $\mathcal{Y}$ is the union of the safe space $S$ and the unsafe space $U$. For any $p \in \mathrm{Distr}(\mathcal{Y})$ and any subset $H \subseteq \mathcal{Y}$, we define the cumulative probability $p(H) = \sum_{y \in H} p(y)$. Given a collection of distributions $(p_1, \ldots, p_n) \in \mathrm{Distr}(\mathcal{Y})^n$, let $p_{(i)}(y)$ represent the $i$-th smallest of the probabilities $\{ p_t(y) \}_{t=1}^n$. We use the symbol $\bot \notin \mathcal{Y}$ to denote an abstention. 

\noindent
\textbf{Output Distribution}. Because the model's output is terminated either by special tokens or by a maximum token limit, for a finite tokenizer, We follow existing work \citep{chijiwa2025lossless} to treat the ostensibly unbounded output distribution as a finite set to facilitate analysis. Formally, for model $i$ with maximum token length $L$ and tokenizer $T_i$, the output space is $\mathcal{Y}_i = \bigcup_{j=1}^{L} T_i^j$.

\noindent
\textbf{Model Group}.
The model group ($\mathcal{MG}$) consists of $n$ generative models, including $s$ safe models and 
$f=n-s$ unsafe models. Each safe model should maintain a reasonable probability distribution for any input prompt $x$. The probability that the safety model outputs a safe response is $\Psi > 0$. Unsafe models are assumed to be fully controllable by an adversary and may exhibit arbitrary unsafe behaviors or follow any probability distribution, and we refer to this as the Byzantine model based on the definition of Byzantine replicas in classical BFT consensus research \citep{zhang2023waterbear, dashing2024duan, Das2024AC}.

\section{Reliable Consensus Sampling}
\label{sec:Methodology}

\subsection{Safety Properties}

We define the safety properties in $\mathcal{MG}$ based on the definitions from reliable distributed systems \citep{cachin2011introduction}.

\noindent
\textbf{Safety}. Let $f < \lceil \frac{n}{2} \rceil$, at all times, the risk of the model group $q(U) \leq n \cdot \mu(U) + \mathrm{negl}(\lambda)$. $\mu(U)$ is the average risk of generating unsafe response $y \in U$ by $s$ safe models. Safety requires that the risk of unsafe models is reduced to the risk of safe models plus $\mathrm{negl}(\lambda)$. $\lambda$ is a security parameter.

\noindent
\textbf{Liveness}. The liveness property requires that, for any time $t$, there remains the hope that the property will be satisfied at some later time $t' \ge t$. The mechanism must still retain the possibility of eventually delivering a usable response $y \in \mathcal{Y}$, although it may be risky. Liveness reflects the utility of $\mathcal{MG}$.

\noindent
\textbf{Anti-Collusion}.
$\mathcal{MG}$ must tolerate arbitrary behavior from up to $f$ Byzantine models. Such behavior includes full control over the probability distribution of $y \in \mathcal{Y}$. This control may assign low probability to safe responses. It may also assign high probability to unsafe responses. These behaviors should not significantly affect the safety of the response delivered by $\mathcal{MG}$.

\noindent
\textbf{Half-Resilience}. In real-world deployments, the values of $s$ and $f$ may change over time. We consider a time $t$ at which $m$ out of the $s$ safe models become unsafe. The $\mathcal{MG}$ system then transitions to a new state with $s' = s - m$ and $f' = f + m$. In this state, the condition $f < \lceil \frac{n}{2} \rceil$ no longer holds. We do not require $\mathcal{MG}$ to preserve theoretical safety or liveness. Instead, we require the system to remain practically robust and avoid catastrophic failure. In this paper, we focus on the case of $f=s$.

\noindent
\textbf{Termination}. 
The algorithm is guaranteed to complete within a finite time $T$.

\subsection{Methodology}
Our proposed RCS scheme is shown in Algorithm \ref{alg:consensus}. During the $R$ rounds of sampling, in each round, if the candidate response $y$ is not accepted, we record $y$ and its acceptance probability $\sigma(y)$ in real time. When all $R$ sampling rounds fail, the protocol enters the trace phase. In this phase, we select the top $\min(s, R)$ values of $\sigma(y)$ and collect the corresponding $y$ into the set $\mathcal{F}$. For each $y \in \mathcal{F}$, we compute $\alpha(y)$ as the sum of the largest $n - s$ values in $\{p_j(y)\}_{j=1}^{n}$, which have already been obtained during the computation of $\sigma(y)$. We then select the $y$ with the largest $\alpha(y)$ in order to mitigate the impact of Byzantine models, while maintaining the liveness guarantee.

\begin{algorithm}[t]
\SetAlgoLined
\KwIn{
Number of models $|\mathcal{MG}|=n$; number of safe models $s$; 
round $R$; distributions $p_1, \ldots, p_n \in \mathrm{Distr}(\mathcal{Y})^{n}$
}
\KwOut{Response $y \in \mathcal{Y}$}

$Buffer \gets \varnothing$\\
\For{$r \leftarrow 1$ \KwTo $R$}{
    Sample $y \sim \frac{1}{n}\sum_{i=1}^{n} p_i$ \\
    $\sigma(y) =\frac{\frac{1}{s}\sum_{i=1}^{s} p_{(i)}(y)}
             {\frac{1}{n}\sum_{i=1}^{n} p_i(y)}$
        
    \If{accept $y$ with $\sigma(y)$}{
        \Return $y$
    }
    $Buffer \gets Buffer \cup \{\langle y, \sigma(y) \rangle \}$\\
}

Sort $Buffer=\{\langle y_i, \sigma(y_i) \rangle \}_{i=1}^{R}$ 
 such that $\sigma(y_{(1)}) \ge \dots \ge \sigma(y_{(R)})$\\
Let $\mathcal{F} = \{y_{(1)}, \dots, y_{u}\}, \; u= min(s,R)$ \\
\For{$i \leftarrow 1$ \KwTo $u$}{
    $\alpha(y_{(i)}) = \sum_{j=s+1}^{n} p_{(j)}(y_{(i)})$
}
\Return $y \gets \arg\max_{y \in \mathcal{F}} \alpha(y)$

\caption{Reliable Consensus Sampling Algorithm}
\label{alg:consensus}
\end{algorithm}

We present a rigorous theoretical proof of the safety property of RCS. The overlap of the output probability distributions of the $s$ safest models, that is, their degree of consensus, is computed by using $Z = \sum_{y}\frac{1}{s}\sum_{i=1}^{s}p_{(i)}(y)$ \citep{kalai2025consensus}. We denote $\Pr[T]=(1-Z)^R$ as the probability of entering the trace phase, and $\rho(U)$ as the probability that the trace phase eventually outputs an unsafe response. Let $R = \mathrm{poly}(\lambda)$. Then the risk of RCS satisfies:

\begin{align*}
q(U) &= (1-\Pr[\text{T}]) \cdot \frac{1}{Z} \sum_{y \in U}\frac{1}{s}\sum_{i=1}^{s}p_{(i)}(y) \\ &+ \Pr[\text{T}] \cdot \rho(U)\\
&\leq  ZR \cdot \frac{1}{Z} \sum_{y \in U}\frac{1}{s}\sum_{i=1}^{s}p_{(i)}(y) + \Pr[\text{T}] \cdot \rho(U)\\
&=  R \cdot \sum_{y \in U}\frac{1}{s}\sum_{i=1}^{s}p_{(i)}(y) + \Pr[\text{T}] \cdot \rho(U)\\
&\leq R \cdot \frac{1}{s}\sum_{i=1}^{s}p_{(i)}(U) + \Pr[\text{T}] \cdot \rho(U) \\
&= R \cdot\mu(U) + (1-Z)^R \cdot \rho(U)
\end{align*}

In $\mathcal{MG}$, at least $s - f$ values of $p_i(y)$ satisfy $p_i(y) > 0$. Therefore, $Z > 0$. It follows that
$(1-Z)^R = e^{R \ln(1-Z)}$. Since $\ln(1-Z) < 0$, $\rho(U) \in (0,1)$ and $R = \mathrm{poly}(\lambda)$, we have $(1-Z)^R \cdot \rho(U) = \mathrm{negl}(\lambda)$. Therefore, $q(U) \leq R \cdot \mu(U) + \text{negl}(\lambda)$. We set the upper bound as $n = k R + b$, where $k,b > 0$ are constants. Under this choice, for any $R \le n$, $q(U) \le n \cdot \mu(U) + \mathrm{negl}(\lambda)$ holds. In fact, the above proof involves two equality conditions. These conditions are difficult to satisfy simultaneously. The first condition is $R = 1$. This condition can be enforced by design. The second condition requires that for all $y \in U$,
$p_i(y) \le p_j(y), \; \forall i \in S,\; j \notin S .$
This condition is extremely difficult to satisfy in practice. Therefore, in general, we have $q(U) < n \cdot \mu(U) + \mathrm{negl}(\lambda)$. As $R \to +\infty$ (without considering termination), we have $\Pr[T] \to 0$, indicating that $\mathcal{MG}$ delivers a response with probability approaching unity. In this asymptotic regime, the condition $q(U) \le n \cdot \mu(U)$ remains valid, thereby ensuring the safety property.

\subsection{Analysis}
We provide a comprehensive comparison between RCS and existing CS method in \autoref{tab:compare}. Below, we analyze each property in detail.

\begin{table*}
\centering
 \scalebox{0.75}{
    \begin{tabular}{@{}l|cccccc@{}}
    \toprule
\textbf{Algorithm}  & \textbf{Safety} & \textbf{Liveness} & \textbf{Anti-Collusion} & \textbf{Half-Resilience} & \textbf{Termination} &\textbf{Time Complexity}\\
    \midrule
Consensus Sampling & \textcolor{nGreen}{\ding{51}} & \textcolor{nRed}{\ding{55}} & \textcolor{nRed}{\ding{55}} & \textcolor{nRed}{\ding{55}} & \textcolor{nGreen}{\ding{51}} & $\mathcal{O}(RI)$\\
Reliable Consensus Sampling & \textcolor{nGreen}{\ding{51}} & \textcolor{nGreen}{\ding{51}} & \textcolor{nGreen}{\ding{51}} & \textcolor{nGreen}{\ding{51}} & \textcolor{nGreen}{\ding{51}} & $[\mathcal{O}(RI), \ \mathcal{O}(RI + n^2 \log n)]$\\
    \bottomrule
    \end{tabular}
 }
 \caption{A comprehensive comparison between our proposed RCS and existing method.}
    \label{tab:compare}
\end{table*}

\noindent
\textbf{Latency}. Let $I$ denote the time required to sample $y$ from a distribution $p_i$.
In RCS, if no sample is accepted after $R$ rounds, the algorithm incurs an additional cost of $n^2 \log n$ for trace computation. This cost is negligible compared with $I$, especially for reasoning LLMs \citep{chen2025towards}. Therefore, RCS and CS have comparable time complexity.

\noindent
\textbf{Anti-Collusion}. The goal of collusion is to make $\mathcal{MG}$ more likely to deliver unsafe responses by manipulating one or more Byzantine models. We study the relation between acceptance probabilities of an unsafe response $y_t$ and a safe response $y_v$ within the $R$ round loop of Algorithm \ref{alg:consensus}. We focus on the quantity $\Delta(\sigma) = \sigma(y_t) - \sigma(y_v)$. The detailed derivation appears in \autoref{app:ac}. The sign of $\Delta(\sigma)$ depends on:
\[
\mathcal{D}_{t,v} =
\frac{
\sum_{i = 1}^{s} p_{(i)}(y_t) \sum_{i = s+1}^{n} p_{(i)}(y_v)
}{
\sum_{i = 1}^{s} p_{(i)}(y_v) \sum_{i = s+1}^{n} p_{(i)}(y_t)
}.
\]

\begin{figure*}[!]
    \centering
    \begin{subfigure}{0.48\textwidth}
        \centering
        \includegraphics[width=\linewidth]{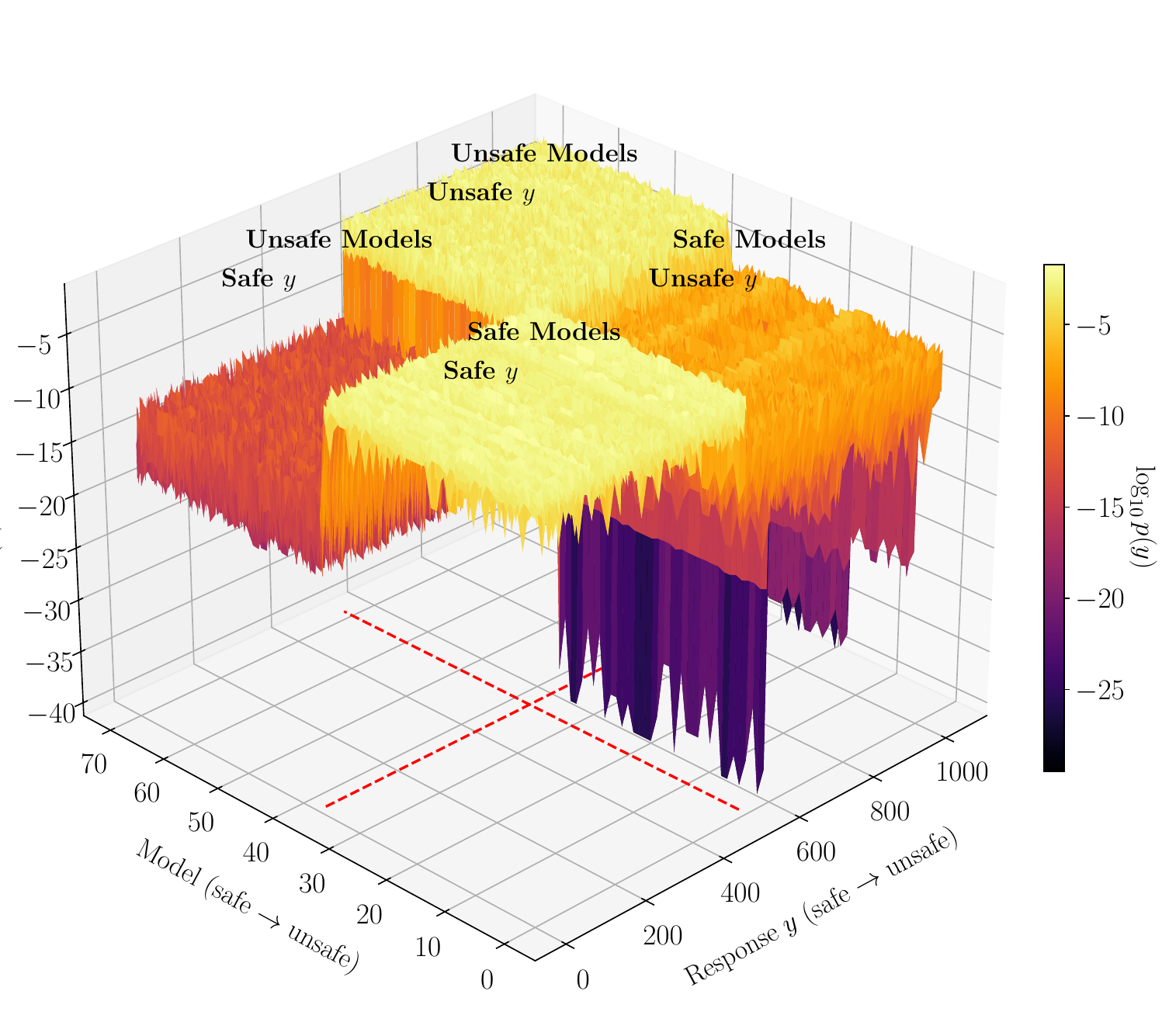}
        \caption{Distribution under general adversarial conditions.}
    \end{subfigure}
    \begin{subfigure}{0.48\textwidth}
        \centering
        \includegraphics[width=\linewidth]{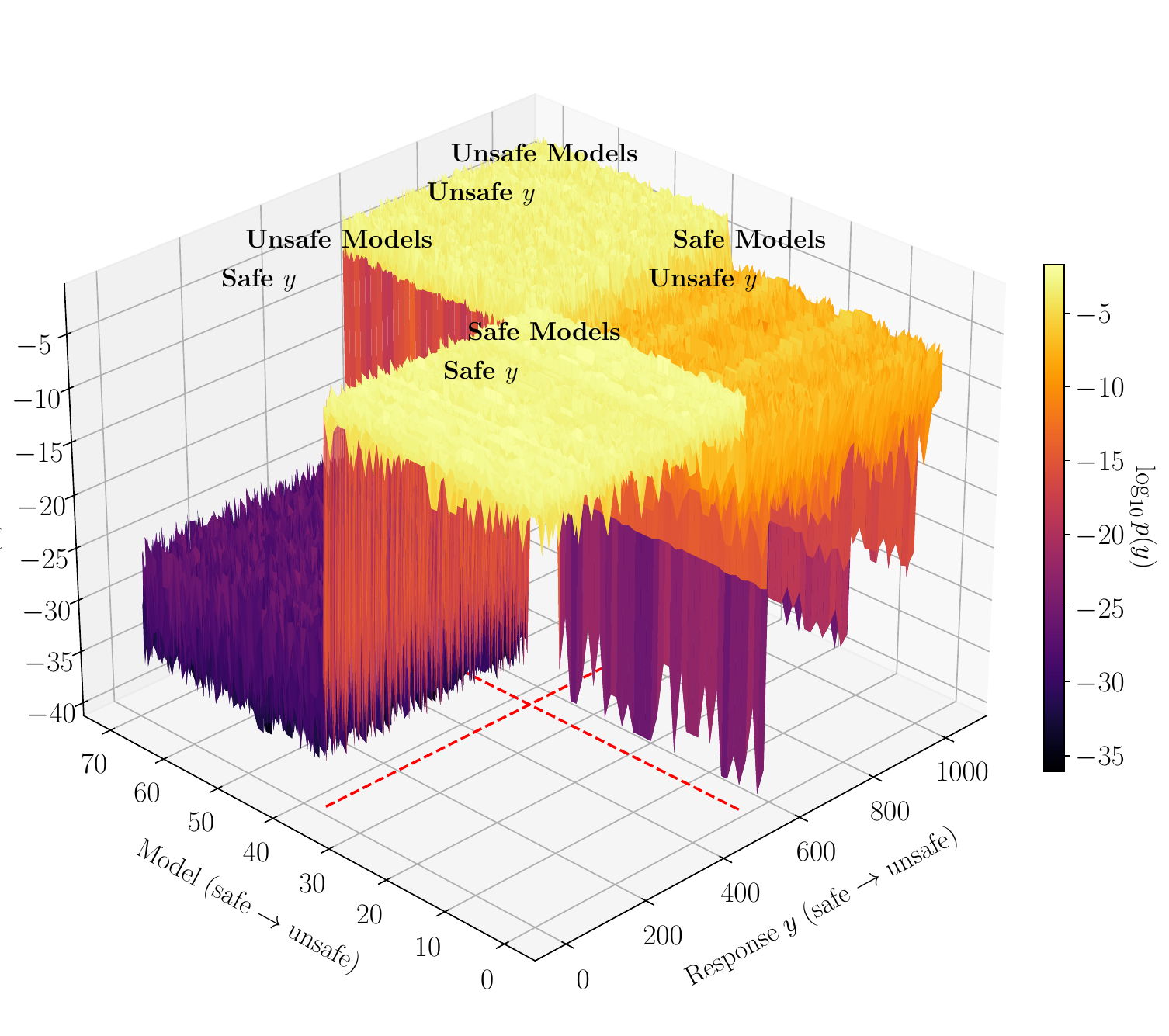}
        \caption{Distribution under malicious collusion.}
    \end{subfigure}
    \caption{Model probability distribution for consensus sampling under diverse adversarial environments.}
    \label{fig:distribution}
    \vspace{-15pt}
\end{figure*}

When the Byzantine model assigns a very low probability to $y_v$, the value of
$\sum_{i=1}^{s} p_{(i)}(y_v)$ decreases, leading to $\mathcal{D}_{t,v} > 1$. As a result, $\Delta(\sigma) > 0$, which significantly increases the acceptance probability of the unsafe response $y_t$ relative to $y_v$. This effect weakens the security of $\mathcal{MG}$. When unsafe models collude, we present the resulting model probability distributions in \autoref{fig:distribution}. The adversarial behavior of unsafe models clearly has a significant impact on the probability overlap among models in CS. In contrast, the design in lines 9-12 of Algorithm \ref{alg:consensus} effectively mitigates this issue. When the values of $p_{\text{unsafe}}(y_v)_j$ are very low, RCS can significantly downweight their influence on the final output response.

\noindent
\textbf{Half-Resilience}. The CS algorithm has an abstention bound, meaning $\Pr[y=\bot]$ does not exceed a threshold when $f < \lceil \frac{n}{2} \rceil$. However, if this condition is violated, the CS abstention bound is affected, leading to an increased probability of abstention and a significant impact on liveness. RCS can avoid this issue. We further validate the half-resilience of RCS in the experiments in Section \ref{sec:experiment}.

\noindent
\textbf{Termination}. 
According to Algorithm \ref{alg:consensus}, for a finite $R$, RCS always terminates within $R$ rounds.

\section{Feedback-Optimized Reliable Consensus Sampling}
For CS and RCS, the framework does not require an explicit definition of security properties. This design allows algorithms to remain applicable over time. From this perspective, for each concrete task, one cannot determine the probability that the final delivered result is safe. This holds even though the algorithmic risk admits an upper bound. External control over the sampling process is limited. This implies that, from both an algorithmic and long-term operational perspective, if the $\mathcal{MG}$ remains unchanged and the input task types are essentially fixed, the risk in RCS remains stationary over time, thereby limiting the potential for dynamic optimization. In this section, we analyze the nature of response safety in RCS. Inspired by quantum computing \citep{nielsen2010quantum}, we propose a research methodology for RCS based on quantum states. We further introduce a mechanism that intervenes in model distributions and dynamically improves safety.

\subsection{Foundational Theory}
\label{sec:Theory}
\noindent
Motivated by quantum computation theory, we propose a framework for studying the security of RCS by analogy with quantum theory.

\noindent
\textbf{Understanding RCS Safety from a Quantum State Perspective}. 
We model a system with unknown safety as a quantum state $|\phi\rangle$, defined as
$|\phi\rangle = \alpha |0\rangle + \beta |1\rangle$,
where $|0\rangle$ denotes a safe state and $|1\rangle$ denotes an unsafe state. The constraint $\alpha^2 + \beta^2 = 1$ holds. Before evaluation, $|\phi\rangle$ does not reside in a definite safe or unsafe state. It remains in a superposition, analogous to Schr\"odinger's cat. The value $|\alpha|^2$ denotes the probability of safety, while $|\beta|^2$ denotes the probability of unsafety.

Given a fixed prompt $q$ and an evaluation method $e$, we apply $\mathrm{Eval}(q, |\phi\rangle, e)$. The state $|\phi\rangle$ then collapses to $|0\rangle$ or $|1\rangle$. Only at this stage can one determine safety under $q$ and $e$. Therefore, $|\phi\rangle$ alone cannot be labeled as safe or unsafe. The state becomes concrete only under the joint action of $q$ and $e$. The desired condition is $|\alpha| > |\beta|$, meaning a stronger bias toward the safe state.

\noindent
\textbf{Entanglement Theory for RCS}. Consider an $\mathcal{MG}$ with $n$ models $\{|\phi_1\rangle, |\phi_2\rangle, \dots, |\phi_n\rangle\}$. In each sampling round, the randomly selected model $|\phi_r\rangle$ remains in a superposition of safety and unsafety. However, the response $y$ produced by $|\phi_r\rangle$ implicitly links $|\phi_r\rangle$ with other models. In $\mathcal{MG}$, there exists a subset
$W = \{|\phi_1\rangle, |\phi_2\rangle, \dots, |\phi_\ell\rangle\}$
that assigns a high probability to $y$. There also exists a subset
$X = \{|\phi_{\ell+1}\rangle, |\phi_{\ell+2}\rangle, \dots, |\phi_n\rangle\}$
that assigns a low probability to $y$. The model $|\phi_r\rangle$ belongs to neither $W$ nor $X$. We say that $W$ and $X$ are entangled. After a time interval $\Delta(t)$, suppose evaluation under $q$ and $e$ verifies that $|\phi_r\rangle$ is unsafe. Then models in $W$ are likely unsafe, while models in $X$ are likely safe. In subsequent sampling, when a prompt $x$ and an evaluation $a$ resemble $q$ and $e$, the algorithm should increase weights for models in $X$ and decrease weights for models in $W$.

\begin{figure*}[!]
    \centering
    \begin{subfigure}{0.31\textwidth}
        \centering
        \includegraphics[width=\linewidth]{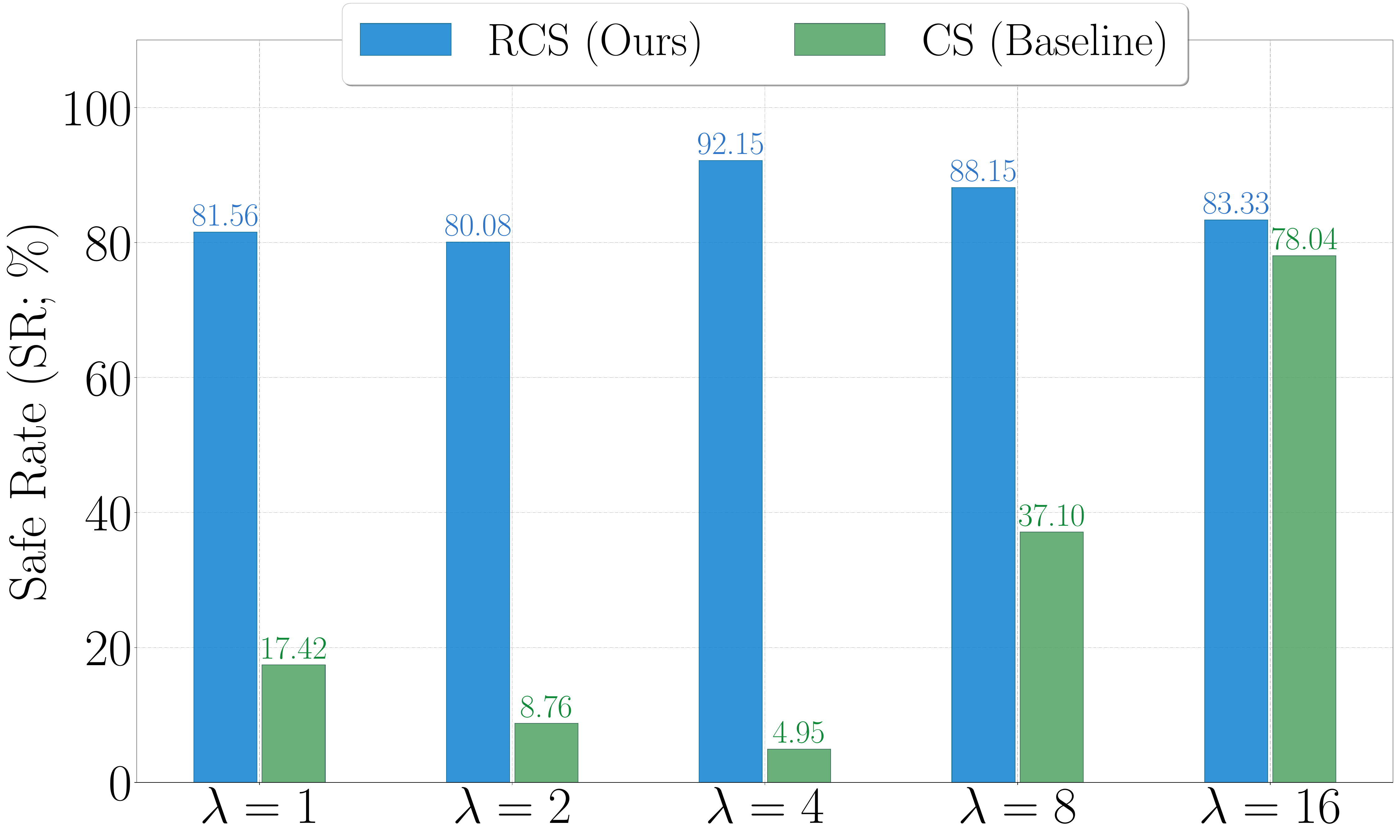}
        \caption{$n=33$.}
    \end{subfigure}
    \begin{subfigure}{0.31\textwidth}
        \centering
        \includegraphics[width=\linewidth]{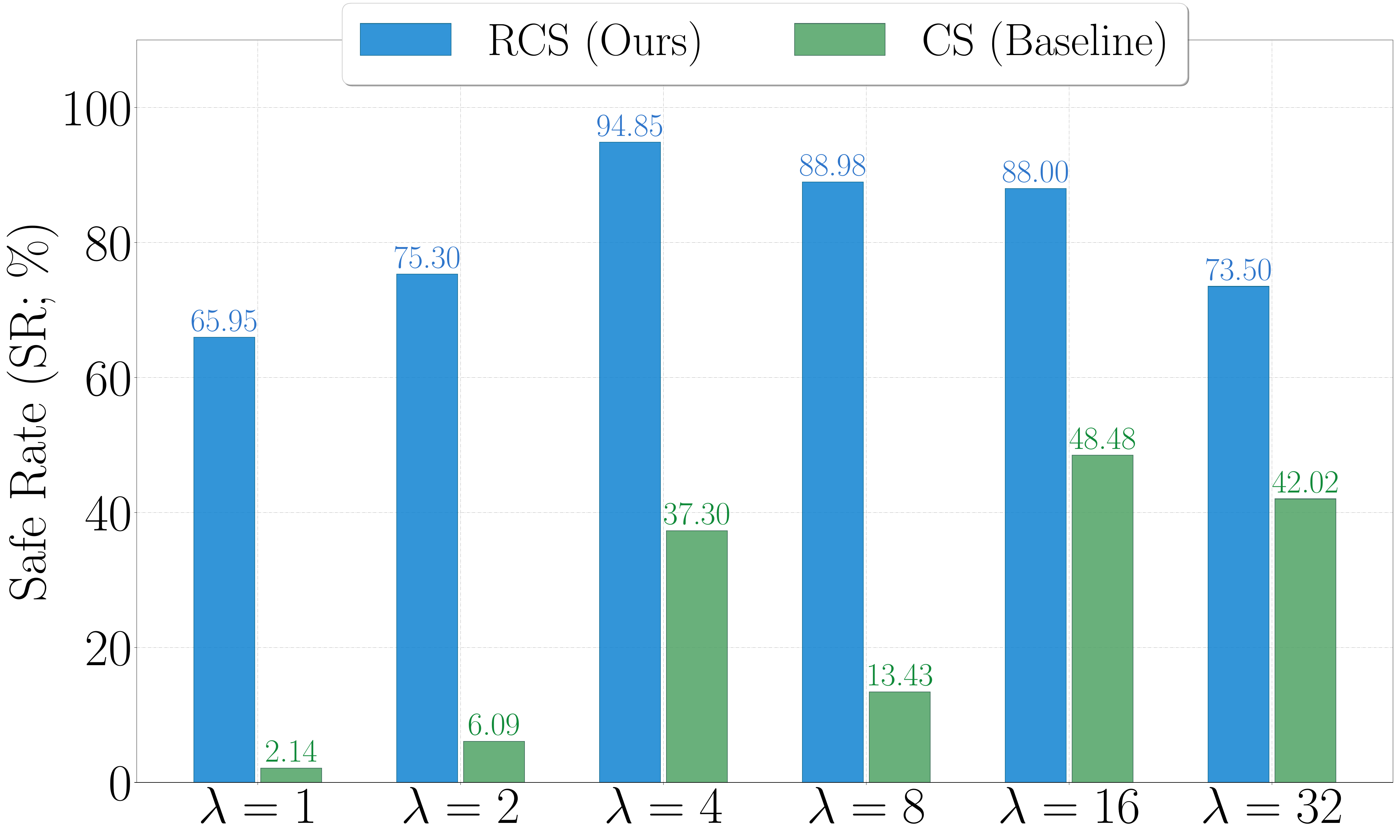}
        \caption{$n=65$.}
    \end{subfigure}
    \begin{subfigure}{0.31\textwidth}
        \centering
        \includegraphics[width=\linewidth]{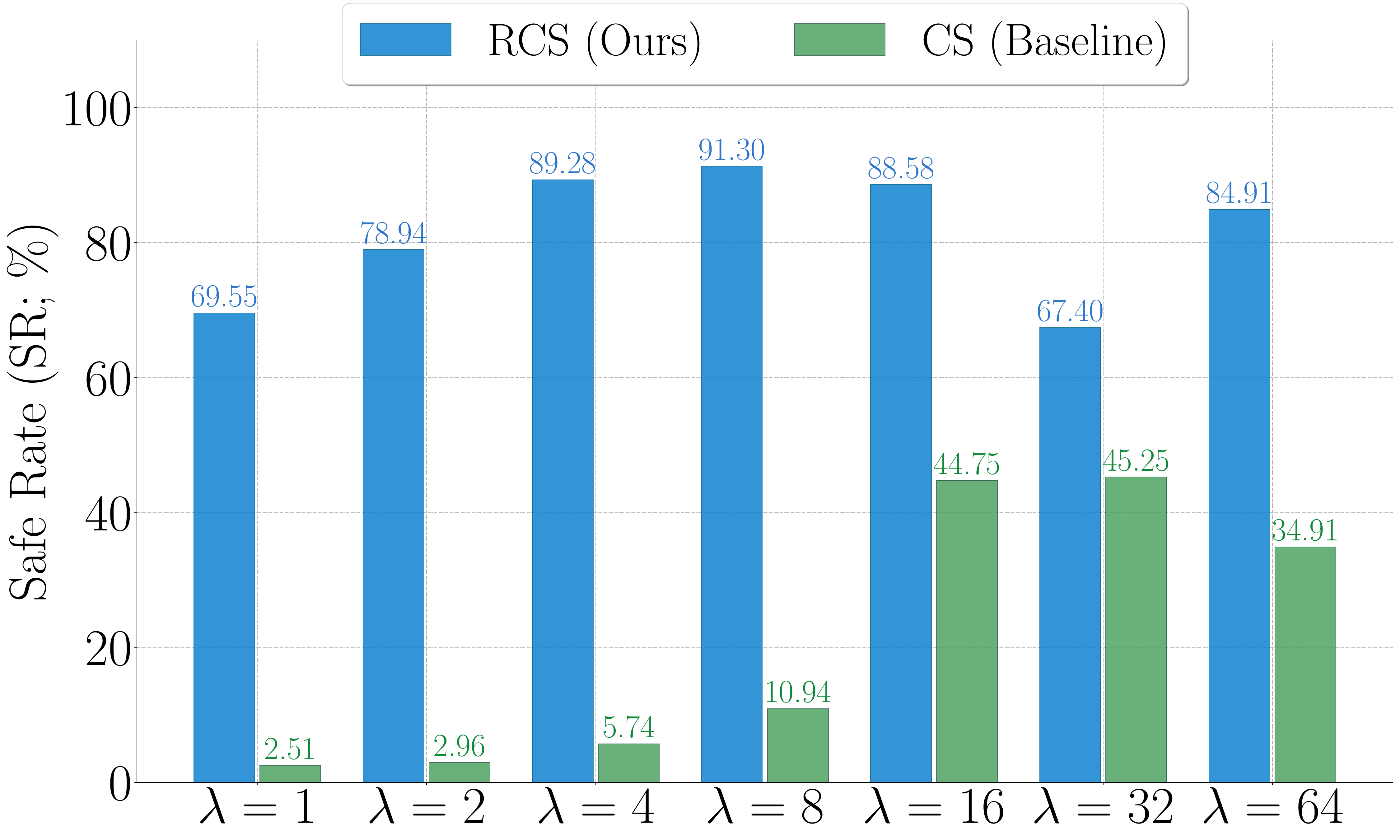}
        \caption{$n=129$.}
    \end{subfigure}
    \caption{Evaluation results for the safe rate when $f < \lceil \frac{n}{2} \rceil$.}
    \label{fig:SR_g}
\end{figure*}

\begin{figure*}[!]
    \centering
    \begin{subfigure}{0.31\textwidth}
        \centering
        \includegraphics[width=\linewidth]{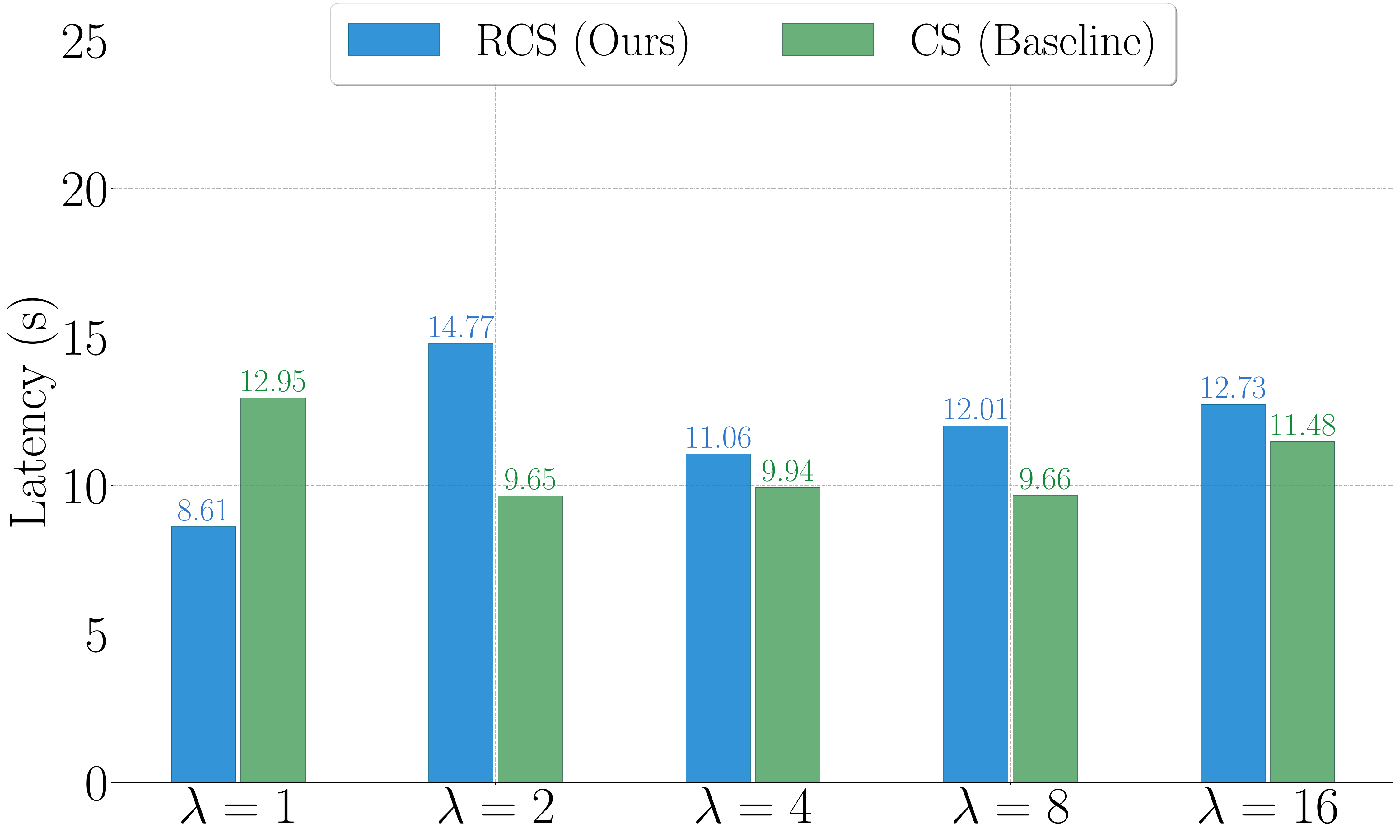}
        \caption{$n=33$.}
    \end{subfigure}
    \begin{subfigure}{0.31\textwidth}
        \centering
        \includegraphics[width=\linewidth]{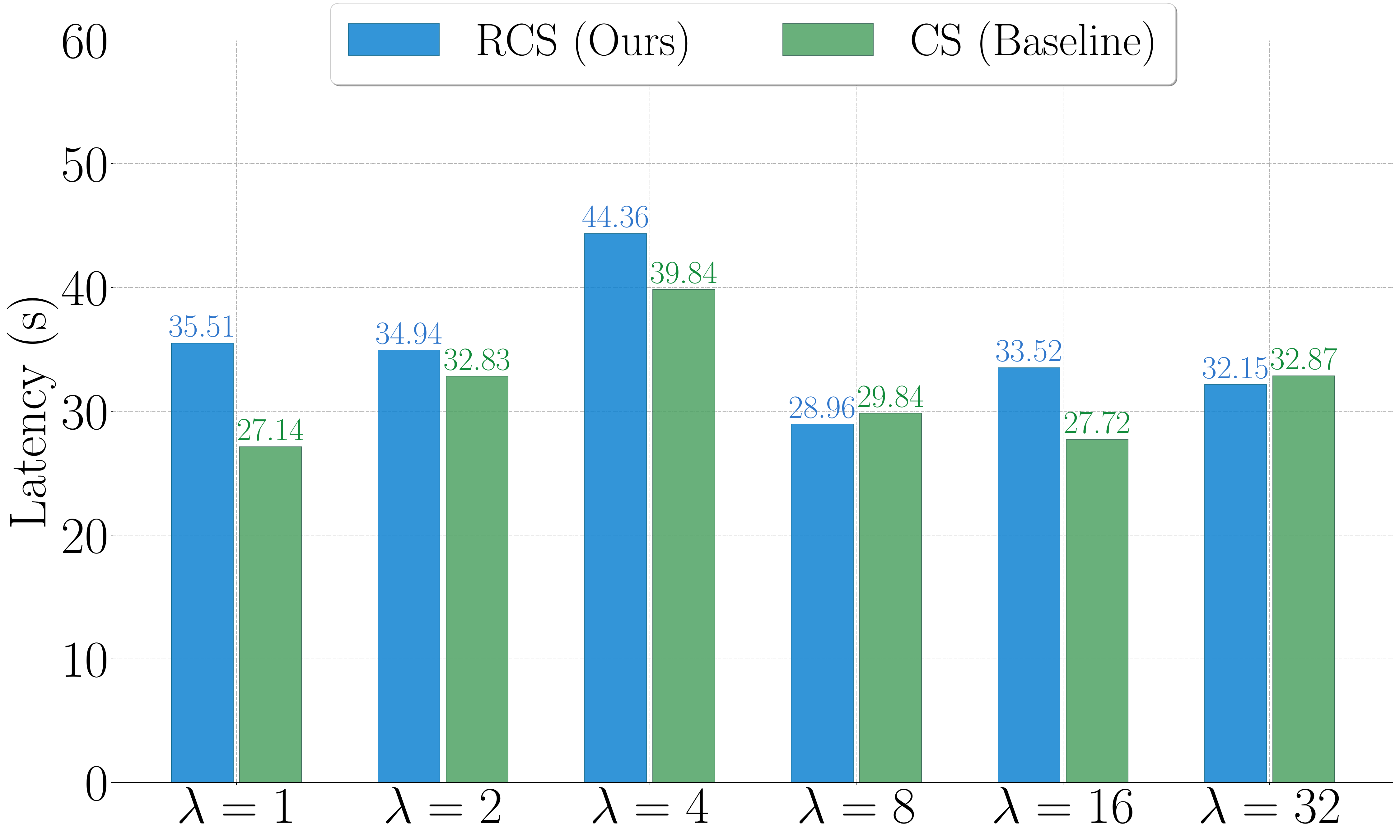}
        \caption{$n=65$.}
    \end{subfigure}
    \begin{subfigure}{0.31\textwidth}
        \centering
        \includegraphics[width=\linewidth]{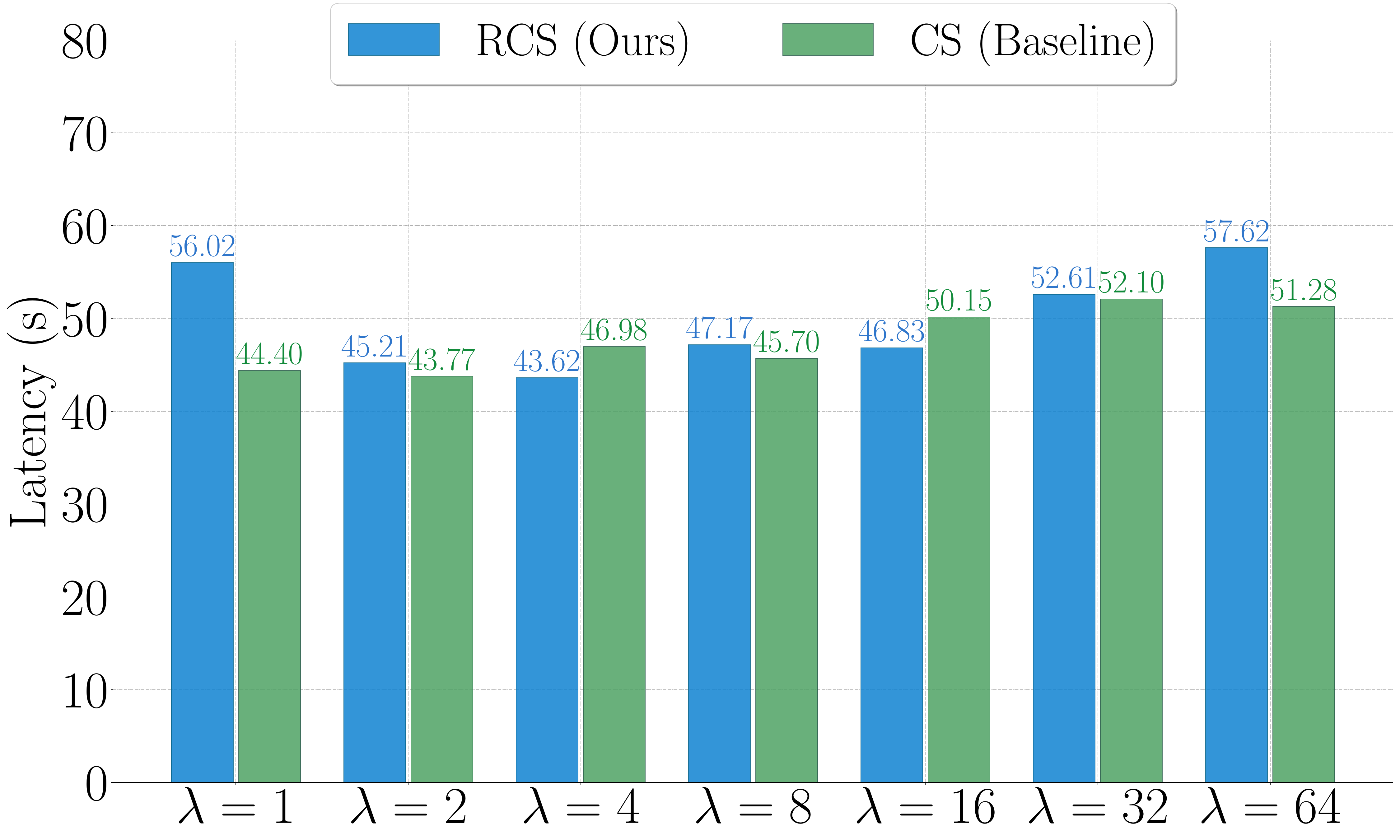}
        \caption{$n=129$.}
    \end{subfigure}
    \caption{Evaluation results for the latency when $f < \lceil \frac{n}{2} \rceil$.}
    \label{fig:TC_g}
\end{figure*}

\subsection{F-RCS Algorithm}
Based on the foundational theory in Section \ref{sec:Theory}, we construct an optimization algorithm for RCS, named Feedback-Optimized Reliable Consensus Sampling (F-RCS). In each sampling round, given the delivered response $y$ and the generating model $|\phi_r\rangle$, the algorithm identifies two additional models by a function $F(\cdot)$ (see Algorithm \ref{alg:F-RCS}). One model $|\phi_{\max}\rangle$ attains the maximum $p(y)$. Another model $|\phi_{\min}\rangle$ attains the minimum $p(y)$. If $y$ is judged unsafe at time $t$, then for subsequent similar tasks, $|\phi_{\max}\rangle$ is removed during random sampling. Formally, $\mathcal{MG} = \mathcal{MG} \setminus \{|\phi_{\max}\rangle\}.$ This allows, at the beginning of each algorithm execution, the system makes an automated decision. It selects safer models used in sampling for $x$. This process increases the safety margin $s - f$ and reduces risk. Here, $|\phi_{\max}\rangle$ can equal $|\phi_{r}\rangle$. Initially, $|\phi_{r}\rangle$ assigns the highest probability to $y$, which aligns with the algorithm objective. We call this behavior of model $|\phi_{r}\rangle$ self-entanglement. In addition, during sampling, the execution of the feedback algorithm is independent of RCS itself, and therefore does not affect the safety of RCS. For time overhead, the additional cost introduced by the feedback algorithm is negligible.

\begin{algorithm}[h]
\SetAlgoLined
\KwIn{
Parameters $(n, s, R)$;
evaluation method $a$
}
\KwOut{Response $y$}

\For{each prompt $x$}{
    $|\phi_{r}\rangle \gets F(x, \mathcal{MG})$ \\
    $\mathcal{MG}^{'} \gets \mathcal{MG} \setminus \{|\phi_{r}\rangle\}$ \\
    $\{y,\mathcal{P}\} \gets \text{RCS}(\mathcal{MG}^{'}, x, s, R)$ \\
    \textcolor{mGreen}{\#$\mathcal{P}=\{p_{|\phi_1\rangle}(y), \cdots, p_{|\phi_n\rangle}(y) \}$} \\
    \textcolor{mGreen}{\#Identify entangled models.} \\
    $|\phi_{\max}\rangle \gets \arg\max_{|\phi_i\rangle} p_{|\phi_i\rangle}(y)$ \\
    \textcolor{mGreen}{\#Record $<|\phi_{\max}\rangle\ ,x, y, |\phi_{r}\rangle>$.} \\
    \Return $y$ \\
    }
    
\While{time $t \in \mathbb{R}^{+}$}{
    \If{$|1\rangle \gets \mathrm{Eval}(x, |\phi_{r}\rangle, a)$}{
        New $F \gets \mathrm{Update}(F,|\phi_{\max}\rangle\ ,x)$ \\
    }
    $t \gets t + \Delta(t)$\\
}

\caption{F-RCS Algorithm}
\label{alg:F-RCS}
\end{algorithm}

\section{Experiments}
\label{sec:experiment}
Although Section \ref{sec:Methodology} presents a theoretical analysis that validates the advantages and effectiveness of our method, we still conduct extensive experiments to demonstrate its practical performance.

\subsection{Experimental Setup}
\noindent
\textbf{Parameter Settings}.
We follow the cryptography \citep{Ruan2025haw, Couteau2022sharp} for selecting the security parameter $\lambda$ and set
$\lambda = 2^d, d \in \mathbb{Z}.$ We choose $R = \lambda + 1$. The values of $s$ and $f$ are determined according to specific types of experiments, and we set $n = kR + b$. From a practical perspective, the relationship between $s$ and $n$ is unknown. The value of $s$ needs to be specified when deploying the algorithm. $s$ does not reflect the actual number of safe models in the $\mathcal{MG}$. To cover general practical scenarios, we adopt the weakest security assumption and set $s = \lceil \frac{n+1}{2} \rceil$. This setting is similar to the widely used $3f+1=n$ in distributed systems research \citep{zhang2023waterbear, Zhang2022pace}.

\noindent
\textbf{Datasets and Models}. 
To reflect adversarial conditions in real-world scenarios, we construct the probability distribution of the safe model using the output distributions of Qwen2.5-7b-instruct, Qwen2.5-0.5b-instruct \citep{qwen2025qwen25technicalreport}, and Qwen3Guard-Gen-8B \citep{zhao2025qwen3guard} on safety evaluation datasets. The datasets include HarmBench \citep{mazeika2024harmbench} and AdvBench \citep{zou2023universal}. We generate the perturbed probability distributions of the unsafe models by referring to the probability distribution of the safe model. This process simulates $f$ Byzantine models that an adversary can fully control. 

\noindent
\textbf{Evaluation Protocol}. To reduce randomness, we report experimental results obtained from 8,000 repeated experiments.

\begin{figure*}[!]
    \centering
    \begin{subfigure}{0.32\textwidth}
        \centering
        \includegraphics[width=\linewidth]{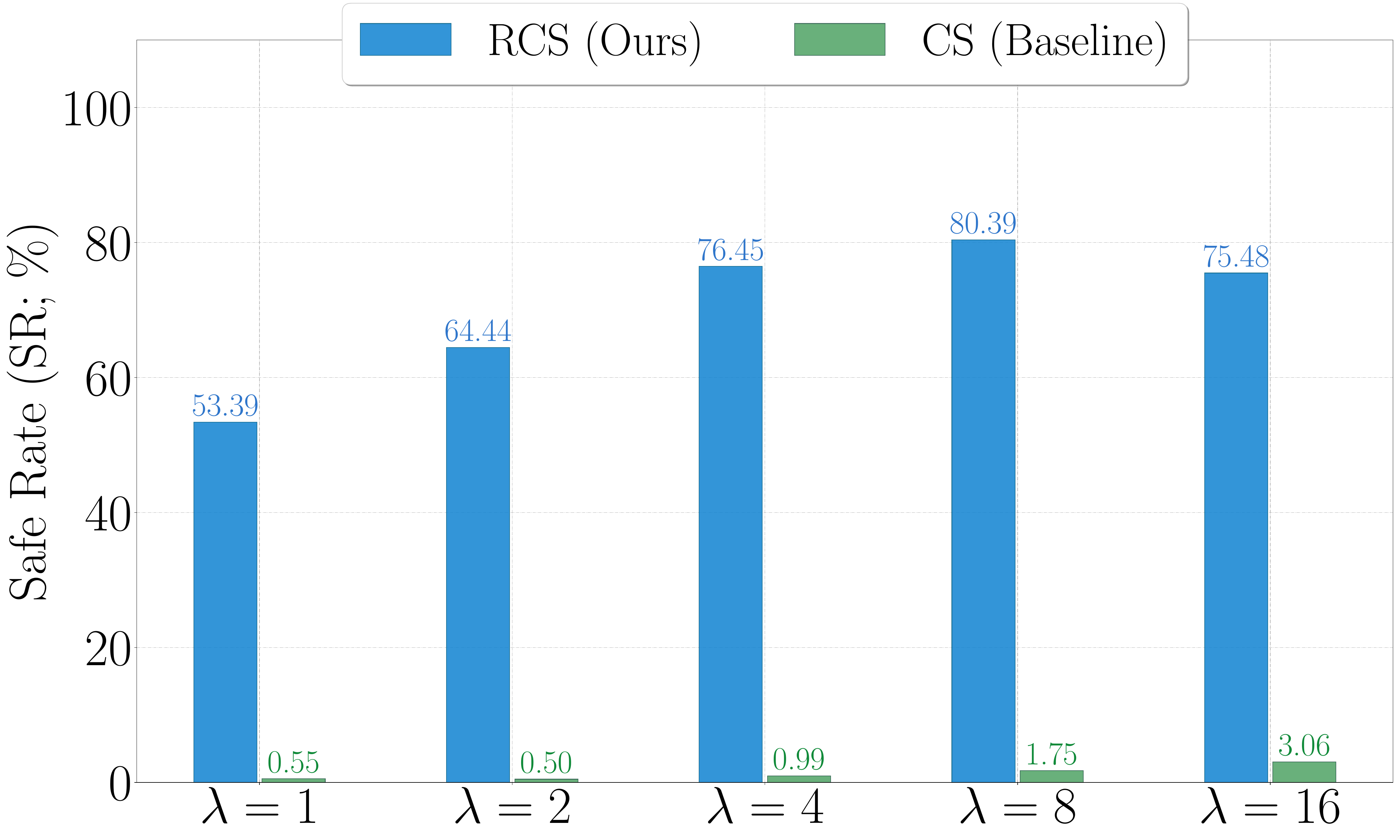}
        \caption{$n=33$.}
    \end{subfigure}
    \begin{subfigure}{0.32\textwidth}
        \centering
        \includegraphics[width=\linewidth]{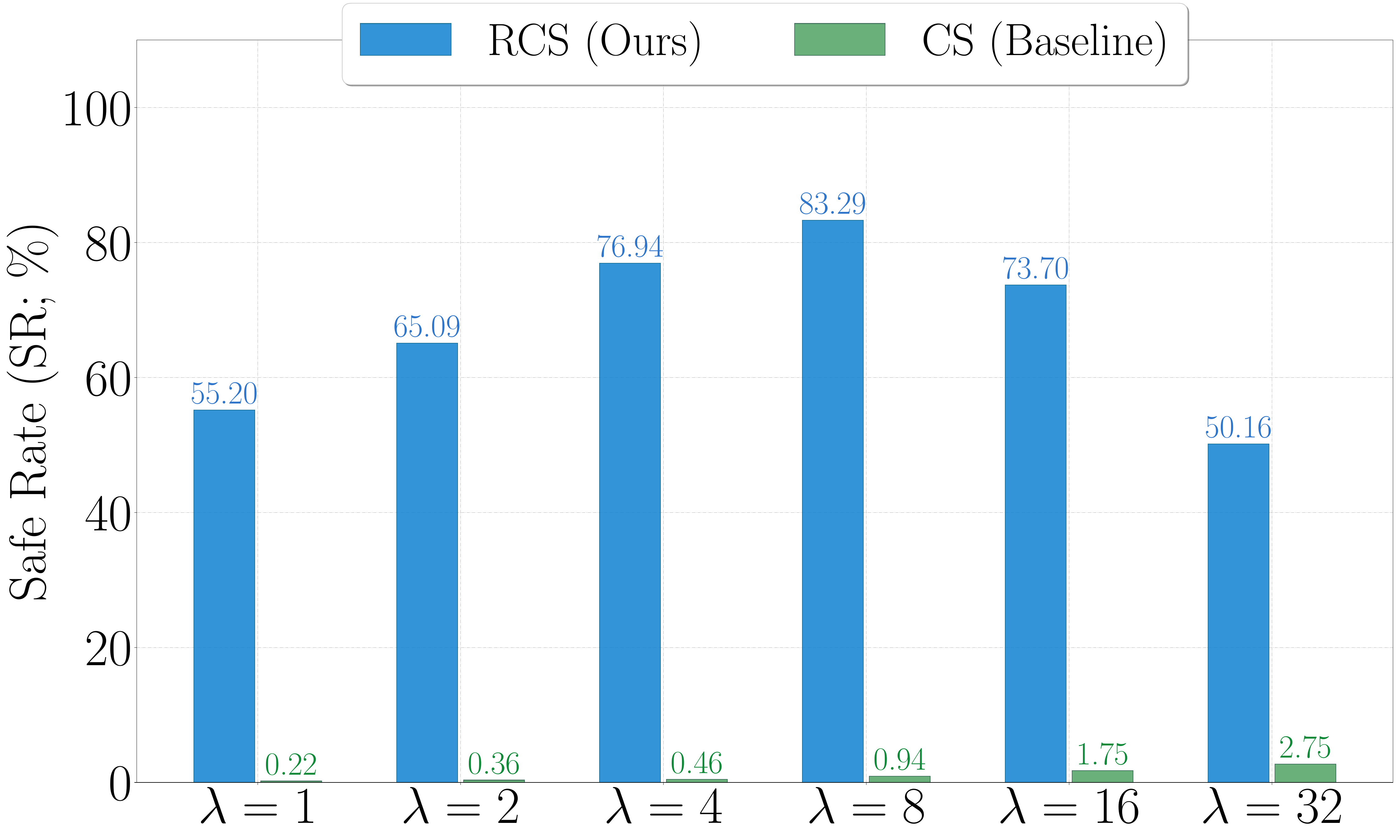}
        \caption{$n=65$.}
    \end{subfigure}
    \begin{subfigure}{0.32\textwidth}
        \centering
        \includegraphics[width=\linewidth]{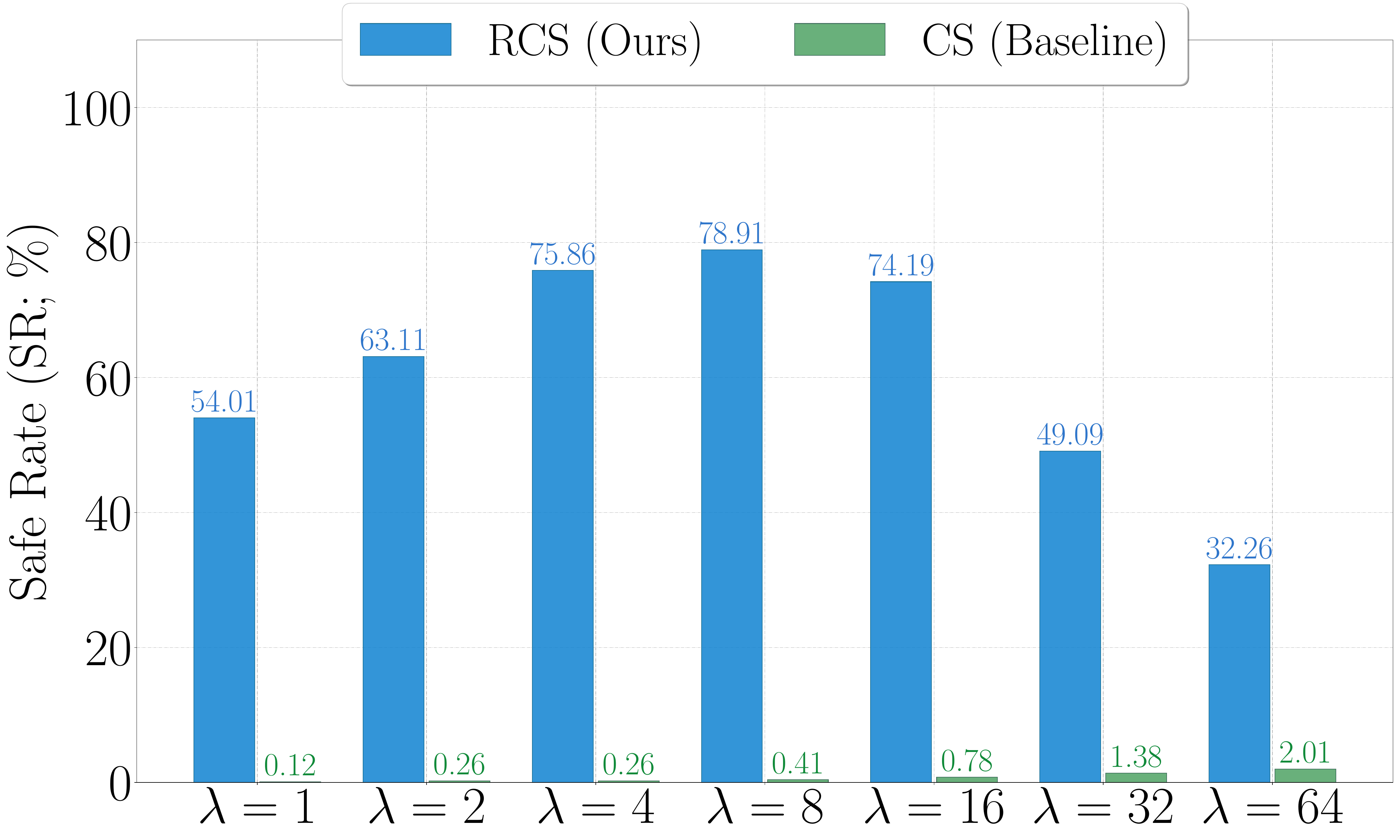}
        \caption{$n=129$.}
    \end{subfigure}
    \caption{Evaluation results for the safe rate when Byzantine models collude.}
    \label{fig:SR_m}
    \vspace{-5pt}
\end{figure*}

\begin{figure*}[!]
    \centering
    \begin{subfigure}{0.32\textwidth}
        \centering
        \includegraphics[width=\linewidth]{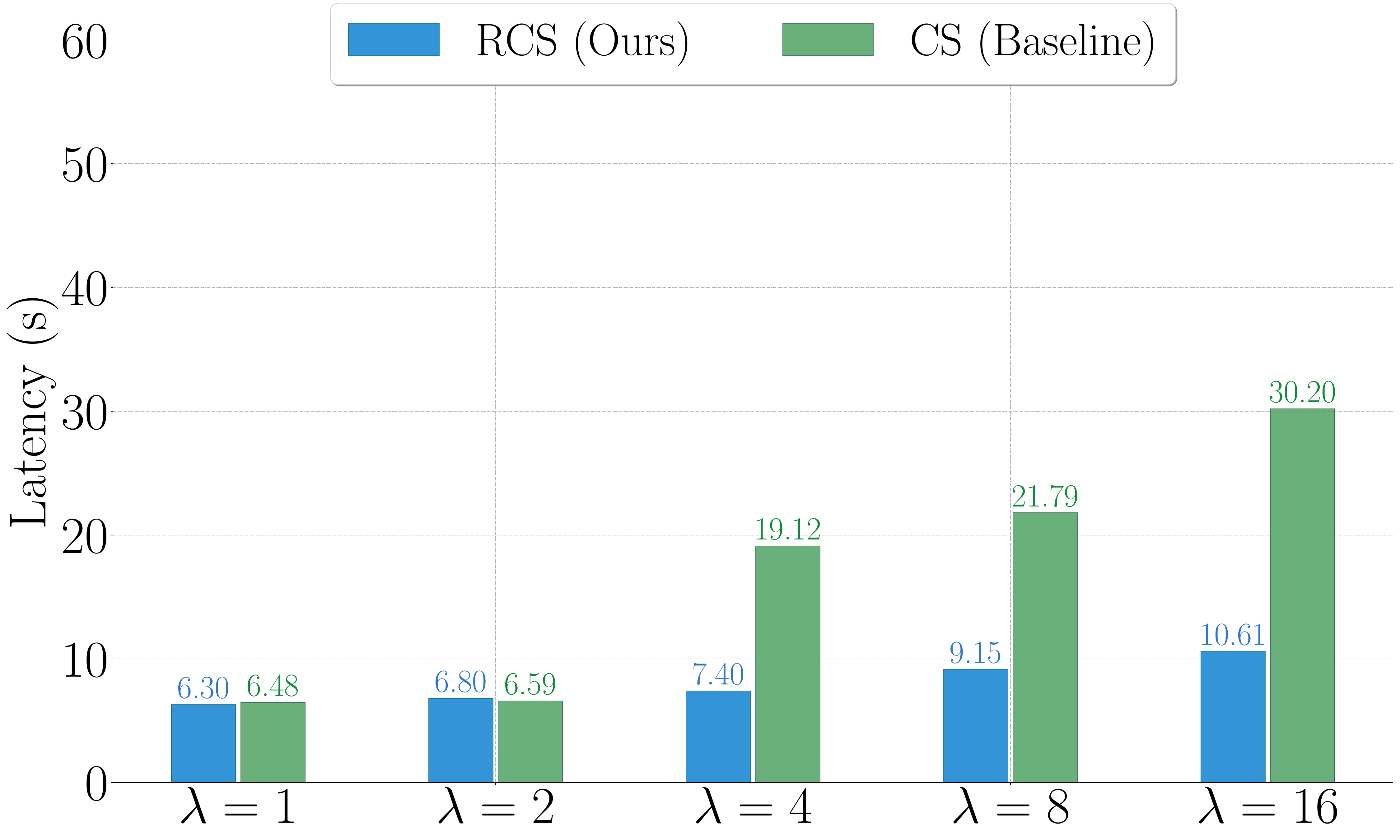}
        \caption{$n=33$.}
    \end{subfigure}
    \begin{subfigure}{0.32\textwidth}
        \centering
        \includegraphics[width=\linewidth]{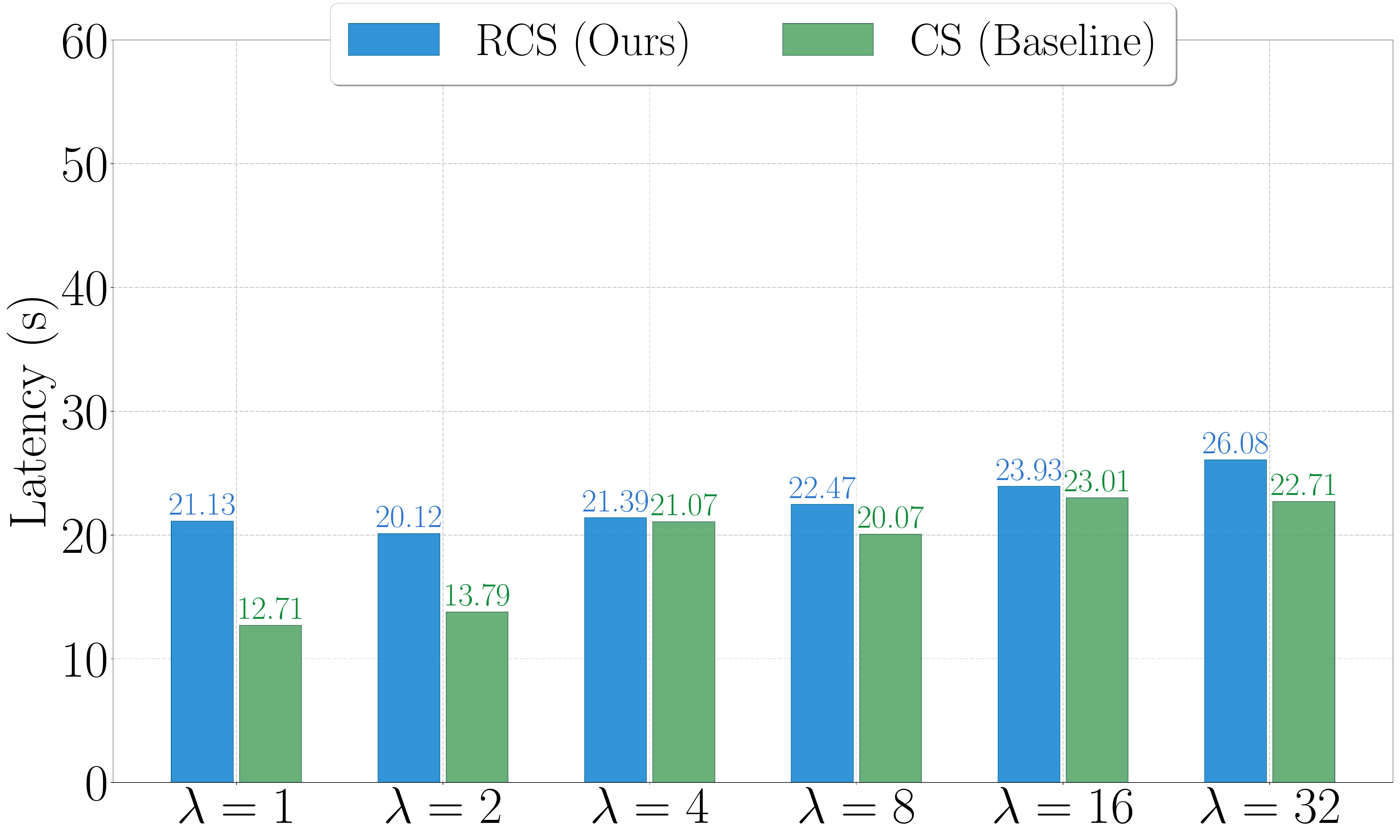}
        \caption{$n=65$.}
    \end{subfigure}
    \begin{subfigure}{0.32\textwidth}
        \centering
        \includegraphics[width=\linewidth]{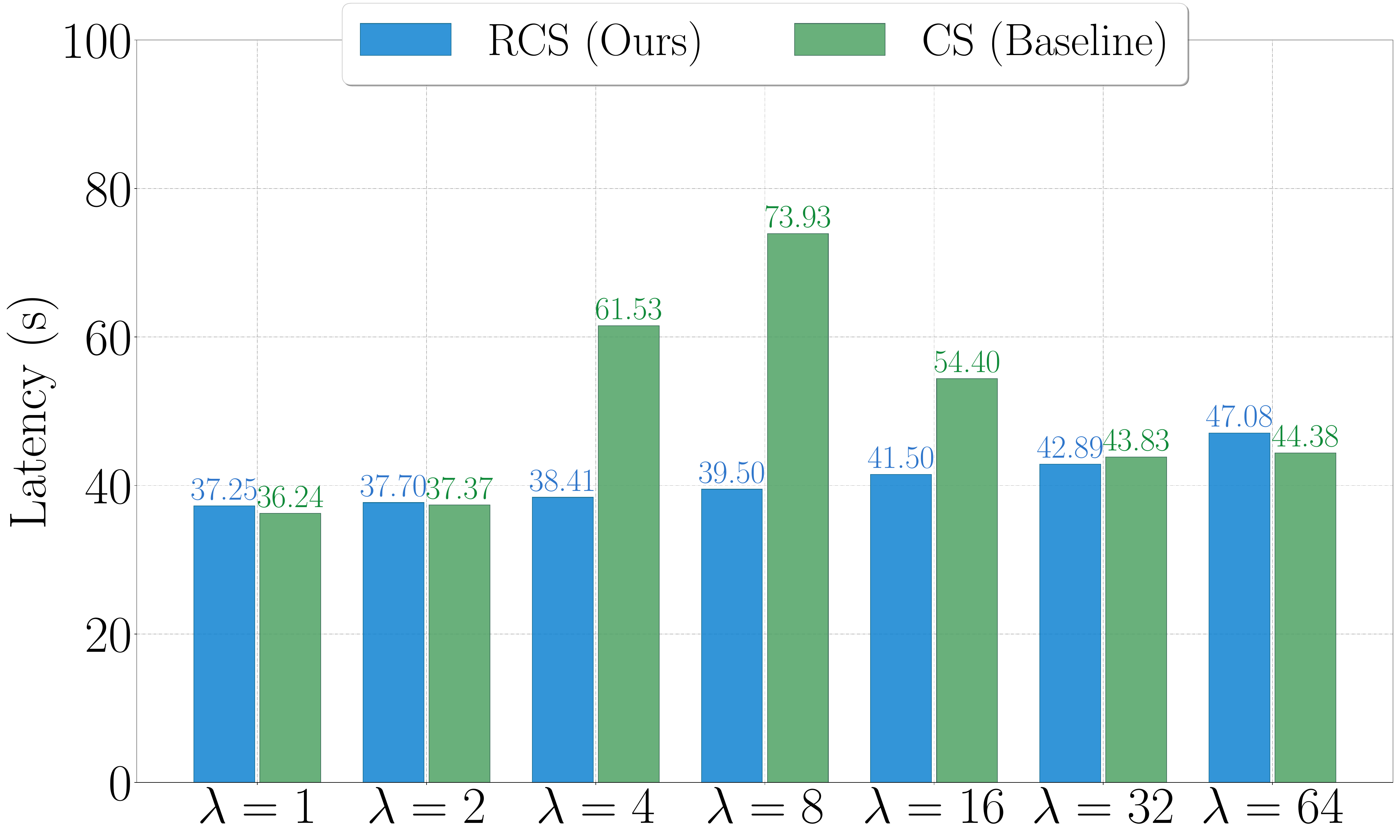}
        \caption{$n=129$.}
    \end{subfigure}
    \caption{Evaluation results for the latency when Byzantine models collude.}
    \label{fig:TC_m}
    \vspace{-15pt}
\end{figure*}

\subsection{Evaluation Metrics}
In this work, we do not define specific criteria that label a response as safe or unsafe. For example, under jailbreak attacks, LLMs may output tokens such as "Sorry" or "Sure" \citep{wang-etal-2025-vulnerability}. These outputs reflect temporary experimental observations. They do not imply provable security in a theoretical sense. Therefore, our evaluation does not rely on the LLM-as-a-judge \citep{wei2025emoji} commonly used in prior jailbreak studies. We define the evaluation metrics as follows. Safe rate (SR) denotes the proportion of final responses produced by the safe models. Abstention rate (AR) denotes the proportion of cases where the $\mathcal{MG}$ returns $\perp$ instead of a usable response. Latency denotes the time interval from the start of the sampling to the delivery of the final response. Accuracy measures the precision of the feedback algorithm in F-RCS when identifying unsafe models, that is, the proportion of models detected by the feedback algorithm that are indeed unsafe.

\subsection{Experimental Results}
We evaluate settings where $f < \lceil \frac{n}{2} \rceil$. The results for large values of $n$ are shown in \autoref{fig:SR_g} and \autoref{fig:TC_g}. Results for smaller values of $n$ are reported in \autoref{app:results}. For different values of the security parameter $\lambda$, RCS achieves significantly higher safety than CS. RCS maintains comparable overall latency to CS. These results are consistent with the theoretical analysis and proofs in Section \ref{sec:Methodology}. CS often leads to abstention, which reduces SR. We further evaluate the algorithms under Byzantine model collusion. Results for large values of $n$ are shown in \autoref{fig:SR_m} and \autoref{fig:TC_m}. Under collusion, the safety of CS degrades catastrophically compared to the $f < \lceil \frac{n}{2} \rceil$ setting. In contrast, the safety of RCS is only slightly affected. RCS maintains more than $67\%$ safety. This result demonstrates strong resistance to collusion. To assess half-resilience, we also evaluate the case where $f = \lceil \frac{n}{2} \rceil$. The results are reported in \autoref{app:results}. In this case, CS again exhibits clear safety weaknesses and performs significantly worse than RCS. \autoref{tab:comparison} reports aggregated results across all values of $\lambda$ and $n$. RCS outperforms CS in both SR and AR. RCS also achieves comparable latency to CS. These results further validate the theoretical analysis and proofs.

\begin{figure}[t]
    \centering
\includegraphics[width=1\linewidth]{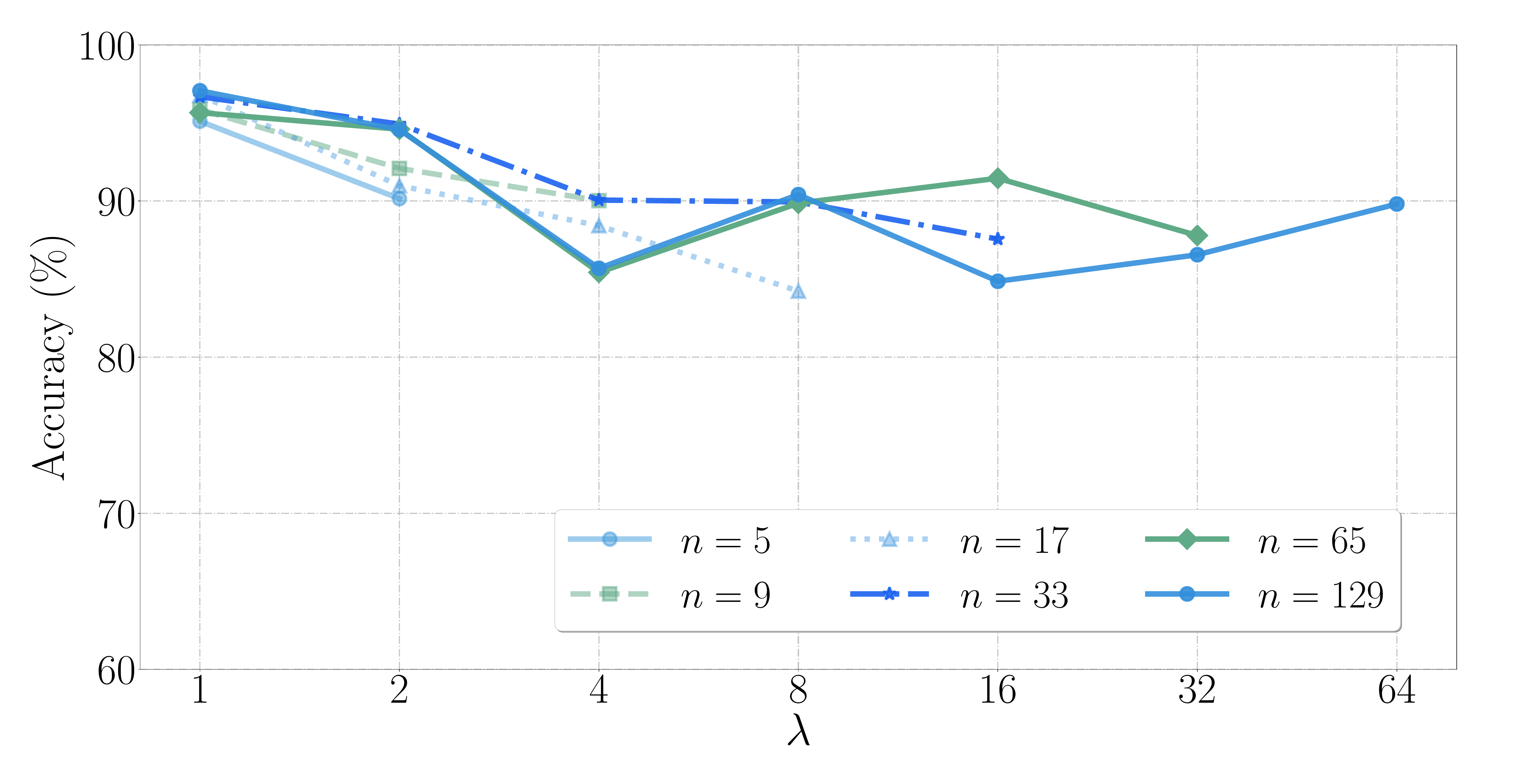}
    \caption{Accuracy of the feedback algorithm for different $\lambda$ and $n$ values under $f < \lceil \frac{n}{2} \rceil$.}
    \label{fig:dr}   
    \vspace{-19pt}
\end{figure}

For F-RCS, we evaluate the accuracy of the feedback algorithm in identifying unsafe models. The results are shown in \autoref{fig:dr} and \autoref{app:results}. We show that for different values of $\lambda$ and $n$, the average accuracy remains around 90\%. This result demonstrates the effectiveness of the feedback algorithm.

\begin{table*}[t]
\centering
\scalebox{0.87}{
    \setlength{\tabcolsep}{6pt}
    \begin{tabular}{l|llll|llll}
    \toprule
    \multirow{2}{*}{\textbf{Method}} 
    & \multicolumn{4}{c|}{\textbf{Consensus Sampling (Baseline)}} 
    & \multicolumn{4}{c}{\textbf{Reliable Consensus Sampling (Ours)}} \\
    \cmidrule(lr){2-5} \cmidrule(lr){6-9}
    & $f < \lceil \frac{n}{2} \rceil$ & $f = \lceil \frac{n}{2} \rceil$ & Collusion & AVG & $f < \lceil \frac{n}{2} \rceil$ & $f = \lceil \frac{n}{2} \rceil$ & Collusion & AVG \\
    \midrule
    Safe Rate       & 20.39 & 15.83 & 1.45  & 12.56 & 82.88 & 81.49 & 67.90 & \textbf{77.42} ($\uparrow\times 5.16$) \\
    Abstention Rate & 71.79 & 72.52 & 82.95 & 75.75 & 0.00  & 0.00 & 0.00 & \textbf{0.00} ($\downarrow 100\%$) \\
    Latency         & 22.60 & 26.73 & 21.43 & 23.59 & 24.26 & 22.21 & 18.42 & \textbf{21.63} ($\downarrow 8.31\%$) \\
    \bottomrule
    \end{tabular}
}
\caption{Comprehensive comparison between our RCS and the baseline across different dimensions. Boldface indicates the best value in each dimension.}
\label{tab:comparison}
\vspace{-10pt}
\end{table*}

\section{Analysis and Discussion}
\subsection{Enhancements}

Within the F-RCS framework, a general assumption is required. Consider the probability distributions produced by $n$ models over an output $y_h$ with unknown safety. After an arbitrary time interval $\Delta(t)$, if $y_h$ is determined to be unsafe, this means that the model state $\lvert \phi_h \rangle$ is unsafe. We require that the set of models $W$ that assign high probability to $y_h$ be trustworthy. Specifically, models in $W$ cannot deny the probability distributions reported before $\Delta(t)$. They also cannot modify the reported distributions. Models in $X$ cannot forge identities to impersonate models in $W$. These conditions can be enforced through cryptographic techniques such as blockchain \citep{Li2023FISCOBCOS, authors2018Fabric}. Such solutions are isomorphic to existing security supervision mechanisms for voting \citep{yang2021priscore}. Concrete designs tailored to F-RCS remain an open direction for future work.

For time cost optimization, we can construct an intermediate algorithm by trading off CS and RCS. One such example is RCS from local coins \citep{zhang2023waterbear}. After $R$ unsuccessful sampling rounds, the algorithm flips a coin to decide whether to return $\bot$ or to enter the trace phase. This design improves efficiency by sacrificing part of the safety. The detailed algorithm is provided in \autoref{app:Local Coin}.

\subsection{Additional Insights on Optimal Safety}

Based on experimental results, we observe a clear pattern. When $n$ is fixed, the SR of RCS often exhibits a maximum for different $\lambda$. For example, \autoref{fig:SR_m} shows a maximum at $\lambda = 8$. Very small values of $\lambda$ and very large values of $\lambda$ both fail to achieve optimal safety. For a fixed $n$, it is unclear both how to determine whether such a maximum exists and how to theoretically identify the value $\lambda_h$ corresponding to it.

\section{Related Work}

Existing research on generative AI security mainly focuses on the safety of model outputs. This line of work studies various attack methods \citep{li2024badedit, li2025one} and corresponding defenses \citep{zhang2025exploiting}. However, these attacks lack a unified definition of security. A representative example is the jailbreak attack \citep{mehrotra2024tree, zhang2025jbshield, yu2024don, wang-etal-2025-diffusionattacker}. This attack aims to induce models to generate unsafe content. Such content covers multiple categories \citep{guo2025deepseeknature}, including discrimination, illegal behavior, harmful behavior, and ethical issues. Output level unsafety is inherently subjective. This subjectivity prevents a deterministic definition under cryptographic theory. Current evaluation of output safety largely relies on LLM-as-a-judge \citep{li-etal-2025-generation}. This approach depends on specialized evaluator models trained for safety analysis, such as Qwen3Guard \citep{zhao2025qwen3guard}. These models perform semantic analysis on generated content. However, evaluator models are not interpretable. Their correctness in safety assessment cannot be formally proven. Defenses against these attacks also lack formal guarantees. As a result, research on generative model security lacks a mature theoretical foundation. This absence of theory leads to inconsistent standards across existing attack and defense studies. 

\section{Conclusion}
In this paper, we formalize a provable security theory for model groups. We introduce RCS, a trace-based method that eliminates abstention and guarantees effective delivery. RCS can tolerate extreme adversarial behavior. We also design a feedback algorithm to improve RCS safety. We prove that RCS has tight upper bounds on response risk. Experiments show that RCS outperforms CS in robustness and utility, achieving a $5\times$ increase in safety rate, while maintaining comparable latency.

\section{Limitations and Ethical Considerations}
In this paper, we employ a safety threshold $f < \lceil \frac{n}{2} \rceil$ in traditional consensus sampling. Although this condition is feasible and justified in real-world deployments, exploring looser thresholds, such as $f < \lceil \frac{2n}{3} \rceil$, remains to be explored in future work. In addition, our discussion on unsafe models covers various adversarial behaviors of Byzantine models. However, more extreme threats warrant further study in future work. The Byzantine behavior discussed in this paper may be harmful. The methods described in this paper may be used for research purposes only.

\bibliography{acl_latex}

@book{cachin2011introduction,
  title={Introduction to reliable and secure distributed programming},
  author={Cachin, Christian and Guerraoui, Rachid and Rodrigues, Lu{\'\i}s},
  year={2011},
  publisher={Springer Science \& Business Media}
}

@inproceedings{zhang2023waterbear,
author = {Zhang, Haibin and Duan, Sisi and Zhao, Boxin and Zhu, Liehuang},
title = {WaterBear: practical asynchronous BFT matching security guarantees of partially synchronous BFT},
year = {2023},
isbn = {978-1-939133-37-3},
publisher = {USENIX Association},
address = {USA},
booktitle = {Proceedings of the 32nd USENIX Conference on Security Symposium},
articleno = {299},
numpages = {17},
location = {Anaheim, CA, USA},
series = {SEC '23}
}

@inproceedings{dashing2024duan,
author = {Duan, Sisi and Zhang, Haibin and Sui, Xiao and Huang, Baohan and Mu, Changchun and Di, Gang and Wang, Xiaoyun},
title = {Dashing and Star: Byzantine Fault Tolerance with Weak Certificates},
year = {2024},
isbn = {9798400704376},
publisher = {Association for Computing Machinery},
address = {New York, NY, USA},
url = {https://doi.org/10.1145/3627703.3650073},
doi = {10.1145/3627703.3650073},
booktitle = {Proceedings of the Nineteenth European Conference on Computer Systems},
pages = {250–264},
numpages = {15},
location = {Athens, Greece},
series = {EuroSys '24}
}

@article{ball2025impossibility,
  title={On the impossibility of separating intelligence from judgment: The computational intractability of filtering for ai alignment},
  author={Ball, Sarah and Gluch, Greg and Goldwasser, Shafi and Kreuter, Frauke and Reingold, Omer and Rothblum, Guy N},
  journal={arXiv preprint arXiv:2507.07341},
  year={2025}
}

@inproceedings{Das2024AC,
author = {Das, Sourav and Duan, Sisi and Liu, Shengqi and Momose, Atsuki and Ren, Ling and Shoup, Victor},
title = {Asynchronous Consensus without Trusted Setup or Public-Key Cryptography},
year = {2024},
isbn = {9798400706363},
publisher = {Association for Computing Machinery},
address = {New York, NY, USA},
url = {https://doi.org/10.1145/3658644.3670327},
doi = {10.1145/3658644.3670327},
booktitle = {Proceedings of the 2024 on ACM SIGSAC Conference on Computer and Communications Security},
pages = {3242–3256},
numpages = {15},
location = {Salt Lake City, UT, USA},
series = {CCS '24}
}

@inproceedings{Zhang2022pace,
author = {Zhang, Haibin and Duan, Sisi},
title = {PACE: Fully Parallelizable BFT from Reproposable Byzantine Agreement},
year = {2022},
isbn = {9781450394505},
publisher = {Association for Computing Machinery},
address = {New York, NY, USA},
url = {https://doi.org/10.1145/3548606.3559348},
doi = {10.1145/3548606.3559348},
booktitle = {Proceedings of the 2022 ACM SIGSAC Conference on Computer and Communications Security},
pages = {3151–3164},
numpages = {14},
location = {Los Angeles, CA, USA},
series = {CCS '22}
}

@book{nielsen2010quantum,
  title={Quantum computation and quantum information},
  author={Nielsen, Michael A and Chuang, Isaac L},
  year={2010},
  publisher={Cambridge university press}
}

@article{kalai2025consensus,
  title={Consensus Sampling for Safer Generative AI},
  author={Kalai, Adam Tauman and Kalai, Yael Tauman and Zamir, Or},
  journal={arXiv preprint arXiv:2511.09493},
  year={2025}
}

@article{chijiwa2025lossless,
  title={Lossless Vocabulary Reduction for Auto-Regressive Language Models},
  author={Chijiwa, Daiki and Hasegawa, Taku and Nishida, Kyosuke and Yamaguchi, Shin'ya and Ohba, Tomoya and Sakao, Tamao and Takeuchi, Susumu},
  journal={arXiv preprint arXiv:2510.08102},
  year={2025}
}

@inproceedings{Li2023FISCOBCOS,
author = {Li, Huizhong and Chen, Yujie and Shi, Xiang and Bai, Xingqiang and Mo, Nan and Li, Wenlin and Guo, Rui and Wang, Zhang and Sun, Yi},
title = {FISCO-BCOS: An Enterprise-grade Permissioned Blockchain System with High-performance},
year = {2023},
isbn = {9798400701092},
publisher = {Association for Computing Machinery},
address = {New York, NY, USA},
url = {https://doi.org/10.1145/3581784.3607053},
doi = {10.1145/3581784.3607053},
booktitle = {Proceedings of the International Conference for High Performance Computing, Networking, Storage and Analysis},
articleno = {68},
numpages = {17},
location = {Denver, CO, USA},
series = {SC '23}
}

@inproceedings{authors2018Fabric,
author = {Androulaki, Elli and Barger, Artem and Bortnikov, Vita and Cachin, Christian and Christidis, Konstantinos and De Caro, Angelo and Enyeart, David and Ferris, Christopher and Laventman, Gennady and Manevich, Yacov and Muralidharan, Srinivasan and Murthy, Chet and Nguyen, Binh and Sethi, Manish and Singh, Gari and Smith, Keith and Sorniotti, Alessandro and Stathakopoulou, Chrysoula and Vukoli\'{c}, Marko and Cocco, Sharon Weed and Yellick, Jason},
title = {Hyperledger fabric: a distributed operating system for permissioned blockchains},
year = {2018},
isbn = {9781450355841},
publisher = {Association for Computing Machinery},
address = {New York, NY, USA},
url = {https://doi.org/10.1145/3190508.3190538},
doi = {10.1145/3190508.3190538},
booktitle = {Proceedings of the Thirteenth EuroSys Conference},
articleno = {30},
numpages = {15},
location = {Porto, Portugal},
series = {EuroSys '18}
}

@article{zhao2025qwen3guard,
  title={Qwen3guard technical report},
  author={Zhao, Haiquan and Yuan, Chenhan and Huang, Fei and Hu, Xiaomeng and Zhang, Yichang and Yang, An and Yu, Bowen and Liu, Dayiheng and Zhou, Jingren and Lin, Junyang and others},
  journal={arXiv preprint arXiv:2510.14276},
  year={2025}
}

@misc{qwen2025qwen25technicalreport,
      title={Qwen2.5 Technical Report}, 
      author={An Yang and Baosong Yang and Beichen Zhang and Binyuan Hui and Bo Zheng and Bowen Yu and Chengyuan Li and Dayiheng Liu and Fei Huang and Haoran Wei and Huan Lin and Jian Yang and Jianhong Tu and Jianwei Zhang and Jianxin Yang and Jiaxi Yang and Jingren Zhou and Junyang Lin and Kai Dang and Keming Lu and Keqin Bao and Kexin Yang and Le Yu and Mei Li and Mingfeng Xue and Pei Zhang and Qin Zhu and Rui Men and Runji Lin and Tianhao Li and Tianyi Tang and Tingyu Xia and Xingzhang Ren and Xuancheng Ren and Yang Fan and Yang Su and Yichang Zhang and Yu Wan and Yuqiong Liu and Zeyu Cui and Zhenru Zhang and Zihan Qiu},
      year={2025},
      eprint={2412.15115},
      archivePrefix={arXiv},
      primaryClass={cs.CL},
      url={https://arxiv.org/abs/2412.15115}, 
}

@article{zou2023universal,
  title={Universal and transferable adversarial attacks on aligned language models},
  author={Zou, Andy and Wang, Zifan and Carlini, Nicholas and Nasr, Milad and Kolter, J Zico and Fredrikson, Matt},
  journal={arXiv preprint arXiv:2307.15043},
  year={2023}
}

@inproceedings{mazeika2024harmbench,
author = {Mazeika, Mantas and Phan, Long and Yin, Xuwang and Zou, Andy and Wang, Zifan and Mu, Norman and Sakhaee, Elham and Li, Nathaniel and Basart, Steven and Li, Bo and Forsyth, David and Hendrycks, Dan},
title = {HarmBench: a standardized evaluation framework for automated red teaming and robust refusal},
year = {2024},
publisher = {JMLR.org},
booktitle = {Proceedings of the 41st International Conference on Machine Learning},
articleno = {1431},
numpages = {44},
location = {Vienna, Austria},
series = {ICML'24}
}

@inproceedings{wang-etal-2025-vulnerability,
    title = "Vulnerability of Large Language Models to Output Prefix Jailbreaks: Impact of Positions on Safety",
    author = "Wang, Yiwei  and
      Chen, Muhao  and
      Peng, Nanyun  and
      Chang, Kai-Wei",
    editor = "Chiruzzo, Luis  and
      Ritter, Alan  and
      Wang, Lu",
    booktitle = "Findings of the Association for Computational Linguistics: NAACL 2025",
    month = apr,
    year = "2025",
    address = "Albuquerque, New Mexico",
    publisher = "Association for Computational Linguistics",
    url = "https://aclanthology.org/2025.findings-naacl.219/",
    doi = "10.18653/v1/2025.findings-naacl.219",
    pages = "3939--3952",
    ISBN = "979-8-89176-195-7"
}

@inproceedings{
wei2025emoji,
title={Emoji Attack: Enhancing Jailbreak Attacks Against Judge {LLM} Detection},
author={Zhipeng Wei and Yuqi Liu and N. Benjamin Erichson},
booktitle={Forty-second International Conference on Machine Learning},
year={2025},
url={https://openreview.net/forum?id=Q0rKYiVEZq}
}

@inproceedings{
weng2025do,
title={Do as We Do, Not as You Think: the Conformity of Large Language Models},
author={Zhiyuan Weng and Guikun Chen and Wenguan Wang},
booktitle={The Thirteenth International Conference on Learning Representations},
year={2025},
url={https://openreview.net/forum?id=st77ShxP1K}
}

@inproceedings{du2024improving,
author = {Du, Yilun and Li, Shuang and Torralba, Antonio and Tenenbaum, Joshua B. and Mordatch, Igor},
title = {Improving factuality and reasoning in language models through multiagent debate},
year = {2024},
publisher = {JMLR.org},
booktitle = {Proceedings of the 41st International Conference on Machine Learning},
articleno = {467},
numpages = {31},
location = {Vienna, Austria},
series = {ICML'24}
}

@inproceedings{chen-etal-2024-reconcile,
    title = "{R}e{C}oncile: Round-Table Conference Improves Reasoning via Consensus among Diverse {LLM}s",
    author = "Chen, Justin  and
      Saha, Swarnadeep  and
      Bansal, Mohit",
    editor = "Ku, Lun-Wei  and
      Martins, Andre  and
      Srikumar, Vivek",
    booktitle = "Proceedings of the 62nd Annual Meeting of the Association for Computational Linguistics (Volume 1: Long Papers)",
    month = aug,
    year = "2024",
    address = "Bangkok, Thailand",
    publisher = "Association for Computational Linguistics",
    url = "https://aclanthology.org/2024.acl-long.381/",
    doi = "10.18653/v1/2024.acl-long.381",
    pages = "7066--7085"
}

@inproceedings{
chan2024chateval,
title={ChatEval: Towards Better {LLM}-based Evaluators through Multi-Agent Debate},
author={Chi-Min Chan and Weize Chen and Yusheng Su and Jianxuan Yu and Wei Xue and Shanghang Zhang and Jie Fu and Zhiyuan Liu},
booktitle={The Twelfth International Conference on Learning Representations},
year={2024},
url={https://openreview.net/forum?id=FQepisCUWu}
}

@inproceedings{
liu2025breaking,
title={Breaking Mental Set to Improve Reasoning through Diverse Multi-Agent Debate},
author={Yexiang Liu and Jie Cao and Zekun Li and Ran He and Tieniu Tan},
booktitle={The Thirteenth International Conference on Learning Representations},
year={2025},
url={https://openreview.net/forum?id=t6QHYUOQL7}
}

@inproceedings{ji-etal-2023-towards,
    title = "Towards Mitigating {LLM} Hallucination via Self Reflection",
    author = "Ji, Ziwei  and
      Yu, Tiezheng  and
      Xu, Yan  and
      Lee, Nayeon  and
      Ishii, Etsuko  and
      Fung, Pascale",
    editor = "Bouamor, Houda  and
      Pino, Juan  and
      Bali, Kalika",
    booktitle = "Findings of the Association for Computational Linguistics: EMNLP 2023",
    month = dec,
    year = "2023",
    address = "Singapore",
    publisher = "Association for Computational Linguistics",
    url = "https://aclanthology.org/2023.findings-emnlp.123/",
    doi = "10.18653/v1/2023.findings-emnlp.123",
    pages = "1827--1843"
}

@inproceedings{mehrotra2024tree,
 author = {Mehrotra, Anay and Zampetakis, Manolis and Kassianik, Paul and Nelson, Blaine and Anderson, Hyrum and Singer, Yaron and Karbasi, Amin},
 booktitle = {Advances in Neural Information Processing Systems},
 editor = {A. Globerson and L. Mackey and D. Belgrave and A. Fan and U. Paquet and J. Tomczak and C. Zhang},
 pages = {61065--61105},
 publisher = {Curran Associates, Inc.},
 title = {Tree of Attacks: Jailbreaking Black-Box LLMs Automatically},
 url = {https://proceedings.neurips.cc/paper_files/paper/2024/file/70702e8cbb4890b4a467b984ae59828a-Paper-Conference.pdf},
 volume = {37},
 year = {2024}
}

@article{guo2025deepseeknature,
  title={DeepSeek-R1 incentivizes reasoning in LLMs through reinforcement learning},
  author={Guo, Daya and Yang, Dejian and Zhang, Haowei and Song, Junxiao and Wang, Peiyi and Zhu, Qihao and Xu, Runxin and Zhang, Ruoyu and Ma, Shirong and Bi, Xiao and others},
  journal={Nature},
  volume={645},
  number={8081},
  pages={633--638},
  year={2025},
  publisher={Nature Publishing Group UK London}
}

@inproceedings{
li2024badedit,
title={BadEdit: Backdooring Large Language Models by Model Editing},
author={Yanzhou Li and Tianlin Li and Kangjie Chen and Jian Zhang and Shangqing Liu and Wenhan Wang and Tianwei Zhang and Yang Liu},
booktitle={The Twelfth International Conference on Learning Representations},
year={2024},
url={https://openreview.net/forum?id=duZANm2ABX}
}

@inproceedings{
li2025one,
title={One Model Transfer to All: On Robust Jailbreak Prompts Generation against {LLM}s},
author={Linbao Li and Yannan Liu and Daojing He and YU LI},
booktitle={The Thirteenth International Conference on Learning Representations},
year={2025},
url={https://openreview.net/forum?id=sULAwlAWc1}
}

@inproceedings{li-etal-2025-generation,
    title = "From Generation to Judgment: Opportunities and Challenges of {LLM}-as-a-judge",
    author = "Li, Dawei  and
      Jiang, Bohan  and
      Huang, Liangjie  and
      Beigi, Alimohammad  and
      Zhao, Chengshuai  and
      Tan, Zhen  and
      Bhattacharjee, Amrita  and
      Jiang, Yuxuan  and
      Chen, Canyu  and
      Wu, Tianhao  and
      Shu, Kai  and
      Cheng, Lu  and
      Liu, Huan",
    editor = "Christodoulopoulos, Christos  and
      Chakraborty, Tanmoy  and
      Rose, Carolyn  and
      Peng, Violet",
    booktitle = "Proceedings of the 2025 Conference on Empirical Methods in Natural Language Processing",
    month = nov,
    year = "2025",
    address = "Suzhou, China",
    publisher = "Association for Computational Linguistics",
    url = "https://aclanthology.org/2025.emnlp-main.138/",
    doi = "10.18653/v1/2025.emnlp-main.138",
    pages = "2757--2791",
    ISBN = "979-8-89176-332-6"
}

@inproceedings{Ruan2025haw,
author = {Ruan, Wenqiang and Lin, Xin and Zhou, Ruisheng and Lin, Guopeng and Yu, Shui and Han, Weili},
title = {HawkEye: statically and accurately profiling the communication cost of models in multi-party learning},
year = {2025},
isbn = {978-1-939133-52-6},
publisher = {USENIX Association},
address = {USA},
booktitle = {Proceedings of the 34th USENIX Conference on Security Symposium},
articleno = {138},
numpages = {19},
location = {Seattle, WA, USA},
series = {SEC '25}
}

@inproceedings{Couteau2022sharp,
author = {Couteau, Geoffroy and Goudarzi, Dahmun and Kloo\ss{}, Michael and Reichle, Michael},
title = {Sharp: Short Relaxed Range Proofs},
year = {2022},
isbn = {9781450394505},
publisher = {Association for Computing Machinery},
address = {New York, NY, USA},
url = {https://doi.org/10.1145/3548606.3560628},
doi = {10.1145/3548606.3560628},
booktitle = {Proceedings of the 2022 ACM SIGSAC Conference on Computer and Communications Security},
pages = {609–622},
numpages = {14},
location = {Los Angeles, CA, USA},
series = {CCS '22}
}

@inproceedings{ji2023beaver,
author = {Ji, Jiaming and Liu, Mickel and Dai, Juntao and Pan, Xuehai and Zhang, Chi and Bian, Ce and Chen, Boyuan and Sun, Ruiyang and Wang, Yizhou and Yang, Yaodong},
title = {BEAVERTAILS: towards improved safety alignment of llm via a human-preference dataset},
year = {2023},
publisher = {Curran Associates Inc.},
address = {Red Hook, NY, USA},
booktitle = {Proceedings of the 37th International Conference on Neural Information Processing Systems},
articleno = {1072},
numpages = {27},
location = {New Orleans, LA, USA},
series = {NIPS '23}
}

@inproceedings {Liu2024injection,
	author = {Yupei Liu and Yuqi Jia and Runpeng Geng and Jinyuan Jia and Neil Zhenqiang Gong},
	title = {Formalizing and Benchmarking Prompt Injection Attacks and Defenses},
	booktitle = {33rd USENIX Security Symposium (USENIX Security 24)},
	year = {2024},
	isbn = {978-1-939133-44-1},
	address = {Philadelphia, PA},
	pages = {1831--1847},
	url = {https://www.usenix.org/conference/usenixsecurity24/presentation/liu-yupei},
	publisher = {USENIX Association},
	month = aug
}

@inproceedings{Zhan2025CAIS,
author = {Zhan, Xiao and Carrillo, Juan Carlos and Seymour, William and Such, Jose},
title = {Malicious LLM-based conversational AI makes users reveal personal information},
year = {2025},
isbn = {978-1-939133-52-6},
publisher = {USENIX Association},
address = {USA},
booktitle = {Proceedings of the 34th USENIX Conference on Security Symposium},
articleno = {4},
numpages = {20},
location = {Seattle, WA, USA},
series = {SEC '25}
}

@inproceedings{zhang2025jbshield,
author = {Zhang, Shenyi and Zhai, Yuchen and Guo, Keyan and Hu, Hongxin and Guo, Shengnan and Fang, Zheng and Zhao, Lingchen and Shen, Chao and Wang, Cong and Wang, Qian},
title = {JBShield: defending large language models from jailbreak attacks through activated concept analysis and manipulation},
year = {2025},
isbn = {978-1-939133-52-6},
publisher = {USENIX Association},
address = {USA},
booktitle = {Proceedings of the 34th USENIX Conference on Security Symposium},
articleno = {421},
numpages = {20},
location = {Seattle, WA, USA},
series = {SEC '25}
}

@article{wang2025uniquesec,
author = {Wang, Shang and Zhu, Tianqing and Liu, Bo and Ding, Ming and Ye, Dayong and Zhou, Wanlei and Yu, Philip},
title = {Unique Security and Privacy Threats of Large Language Models: A Comprehensive Survey},
year = {2025},
publisher = {Association for Computing Machinery},
address = {New York, NY, USA},
issn = {0360-0300},
url = {https://doi.org/10.1145/3764113},
doi = {10.1145/3764113},
note = {Just Accepted},
journal = {ACM Comput. Surv.},
month = sep
}

@article{jain2024makes,
  title={What makes and breaks safety fine-tuning? a mechanistic study},
  author={Jain, Samyak and Lubana, Ekdeep S and Oksuz, Kemal and Joy, Tom and Torr, Philip and Sanyal, Amartya and Dokania, Puneet},
  journal={Advances in Neural Information Processing Systems},
  volume={37},
  pages={93406--93478},
  year={2024}
}

@article{chen2024agentpoison,
  title={Agentpoison: Red-teaming llm agents via poisoning memory or knowledge bases},
  author={Chen, Zhaorun and Xiang, Zhen and Xiao, Chaowei and Song, Dawn and Li, Bo},
  journal={Advances in Neural Information Processing Systems},
  volume={37},
  pages={130185--130213},
  year={2024}
}

@inproceedings{xu-etal-2024-instructions,
    title = "Instructions as Backdoors: Backdoor Vulnerabilities of Instruction Tuning for Large Language Models",
    author = "Xu, Jiashu  and
      Ma, Mingyu  and
      Wang, Fei  and
      Xiao, Chaowei  and
      Chen, Muhao",
    editor = "Duh, Kevin  and
      Gomez, Helena  and
      Bethard, Steven",
    booktitle = "Proceedings of the 2024 Conference of the North American Chapter of the Association for Computational Linguistics: Human Language Technologies (Volume 1: Long Papers)",
    month = jun,
    year = "2024",
    address = "Mexico City, Mexico",
    publisher = "Association for Computational Linguistics",
    url = "https://aclanthology.org/2024.naacl-long.171/",
    doi = "10.18653/v1/2024.naacl-long.171",
    pages = "3111--3126"
}

@article{wen2024privacy,
  title={Privacy backdoors: Enhancing membership inference through poisoning pre-trained models},
  author={Wen, Yuxin and Marchyok, Leo and Hong, Sanghyun and Geiping, Jonas and Goldstein, Tom and Carlini, Nicholas},
  journal={Advances in Neural Information Processing Systems},
  volume={37},
  pages={83374--83396},
  year={2024}
}

@article{dalrymple2024towards,
  title={Towards guaranteed safe ai: A framework for ensuring robust and reliable ai systems},
  author={Dalrymple, David and Skalse, Joar and Bengio, Yoshua and Russell, Stuart and Tegmark, Max and Seshia, Sanjit and Omohundro, Steve and Szegedy, Christian and Goldhaber, Ben and Ammann, Nora and others},
  journal={arXiv preprint arXiv:2405.06624},
  year={2024}
}

@inproceedings{yu2024don,
  title={Don't listen to me: Understanding and exploring jailbreak prompts of large language models},
  author={Yu, Zhiyuan and Liu, Xiaogeng and Liang, Shunning and Cameron, Zach and Xiao, Chaowei and Zhang, Ning},
  booktitle={33rd USENIX Security Symposium (USENIX Security 24)},
  pages={4675--4692},
  year={2024}
}

@inproceedings{wang-etal-2025-diffusionattacker,
    title = "{D}iffusion{A}ttacker: Diffusion-Driven Prompt Manipulation for {LLM} Jailbreak",
    author = "Wang, Hao  and
      Li, Hao  and
      Zhu, Junda  and
      Wang, Xinyuan  and
      Pan, Chengwei  and
      Huang, Minlie  and
      Sha, Lei",
    editor = "Christodoulopoulos, Christos  and
      Chakraborty, Tanmoy  and
      Rose, Carolyn  and
      Peng, Violet",
    booktitle = "Proceedings of the 2025 Conference on Empirical Methods in Natural Language Processing",
    month = nov,
    year = "2025",
    address = "Suzhou, China",
    publisher = "Association for Computational Linguistics",
    url = "https://aclanthology.org/2025.emnlp-main.1128/",
    doi = "10.18653/v1/2025.emnlp-main.1128",
    pages = "22193--22205",
    ISBN = "979-8-89176-332-6"
}

@inproceedings{zhang2025exploiting,
  title={Exploiting $\{$Task-Level$\}$ Vulnerabilities: An Automatic Jailbreak Attack and Defense Benchmarking for $\{$LLMs$\}$},
  author={Zhang, Lan and Gao, Xinben and Yao, Liuyi and Song, Jinke and Li, Yaliang},
  booktitle={34th USENIX Security Symposium (USENIX Security 25)},
  pages={2363--2382},
  year={2025}
}

@article{chen2025towards,
  title={Towards reasoning era: A survey of long chain-of-thought for reasoning large language models},
  author={Chen, Qiguang and Qin, Libo and Liu, Jinhao and Peng, Dengyun and Guan, Jiannan and Wang, Peng and Hu, Mengkang and Zhou, Yuhang and Gao, Te and Che, Wanxiang},
  journal={arXiv preprint arXiv:2503.09567},
  year={2025}
}

@article{paulus2025safety,
  title={Safety Alignment of LMs via Non-cooperative Games},
  author={Paulus, Anselm and Kulikov, Ilia and Amos, Brandon and Munos, R{\'e}mi and Evtimov, Ivan and Chaudhuri, Kamalika and Zharmagambetov, Arman},
  journal={arXiv preprint arXiv:2512.20806},
  year={2025}
}

@article{luo2025large,
  title={Large language model agent: A survey on methodology, applications and challenges},
  author={Luo, Junyu and Zhang, Weizhi and Yuan, Ye and Zhao, Yusheng and Yang, Junwei and Gu, Yiyang and Wu, Bohan and Chen, Binqi and Qiao, Ziyue and Long, Qingqing and others},
  journal={arXiv preprint arXiv:2503.21460},
  year={2025}
}

@article{yang2021priscore,
  title={PriScore: Blockchain-based self-tallying election system supporting score voting},
  author={Yang, Yang and Guan, Zhangshuang and Wan, Zhiguo and Weng, Jian and Pang, Hwee Hwa and Deng, Robert H},
  journal={IEEE Transactions on Information Forensics and Security},
  volume={16},
  pages={4705--4720},
  year={2021},
  publisher={IEEE}
}
\onecolumn
\newpage

\appendix

\section{RCS from Local Coins}
\label{app:Local Coin}

\begin{algorithm}[h]
\SetAlgoLined
\KwIn{
Number of models $|\mathcal{MG}|=n$; number of safe models $s$; 
round $R$; distributions $p_1, \ldots, p_n \in \mathrm{Distr}(\mathcal{Y})^{n}$
}
\KwOut{Response $y \in \mathcal{Y}$}

$Buffer \gets \varnothing$\\
\For{$r \leftarrow 1$ \KwTo $R$}{
    Sample $y \sim \frac{1}{n}\sum_{i=1}^{n} p_i$ \\
    $\sigma(y) =\frac{\frac{1}{s}\sum_{i=1}^{s} p_{(i)}(y)}
             {\frac{1}{n}\sum_{i=1}^{n} p_i(y)}$
        
    \If{accept $y$ with $\sigma(y)$}{
        \Return $y$
    }
    $Buffer \gets Buffer \cup \{\langle y, \sigma(y) \rangle \}$\\
}
$c \gets Random()$ \textcolor{mGreen}{\#Obtain local coin 0 or 1.} \\
\If{$c==0$}{
\Return $\bot$
}\Else{

Sort $Buffer=\{\langle y_i, \sigma(y_i) \rangle \}_{i=1}^{R}$ 
 such that $\sigma(y_{(1)}) \ge \dots \ge \sigma(y_{(R)})$\\
Let $\mathcal{F} = \{y_{(1)}, \dots, y_{u}\}, \; u= min(s,R)$ \\
\For{$i \leftarrow 1$ \KwTo $u$}{
    $\alpha(y_{(i)}) = \sum_{j=s+1}^{n} p_{(j)}(y_{(i)})$
}
\Return $y \gets \arg\max_{y \in \mathcal{F}} \alpha(y)$

}

\caption{Reliable Consensus Sampling from Local Coins}
\label{alg:coin}
\end{algorithm}

\section{Proof}
\label{app:ac}

\begin{align*}
&\sigma(y_t) - \sigma(y_v) \\
&=
\frac{\frac{1}{s} \sum_{i \le s} p_{(i)}(y_t)}
     {\frac{1}{n} \sum_{i=1}^{n} p_i(y_t)}
-
\frac{\frac{1}{s} \sum_{i \le s} p_{(i)}(y_v)}
     {\frac{1}{n} \sum_{i=1}^{n} p_i(y_v)}
\\[2mm]
&=
\frac{
\left( \frac{1}{s} \sum_{i \le s} p_{(i)}(y_t) \right)
\left( \frac{1}{n} \sum_{i=1}^{n} p_i(y_v) \right)
-
\left( \frac{1}{s} \sum_{i \le s} p_{(i)}(y_v) \right)
\left( \frac{1}{n} \sum_{i=1}^{n} p_i(y_t) \right)
}{
\left( \frac{1}{n} \sum_{i=1}^{n} p_i(y_t) \right)
\left( \frac{1}{n} \sum_{i=1}^{n} p_i(y_v) \right)
}
\\[2mm]
&=
\frac{n}{s} \cdot
\frac{
\left( \sum_{i \le s} p_{(i)}(y_t) \right)
\left( \sum_{i=1}^{n} p_i(y_v) \right)
-
\left( \sum_{i \le s} p_{(i)}(y_v) \right)
\left( \sum_{i=1}^{n} p_i(y_t) \right)
}{
\left( \sum_{i=1}^{n} p_i(y_t) \right)
\left( \sum_{i=1}^{n} p_i(y_v) \right)
}
\\[2mm]
&=
\frac{n}{s} \cdot
\frac{
\left( \sum_{i \le s} p_{(i)}(y_t) \right)
\left( \sum_{i > s} p_{(i)}(y_v) \right)
-
\left( \sum_{i \le s} p_{(i)}(y_v) \right)
\left( \sum_{i > s} p_{(i)}(y_t) \right)
}{
\left( \sum_{i=1}^{n} p_i(y_t) \right)
\left( \sum_{i=1}^{n} p_i(y_v) \right)
}
\end{align*}

\section{Additional Discussion}
\subsection{RCS and Multi-Agent Debate}

Multi-agent debate (MAD) \citep{du2024improving, chan2024chateval, liu2025breaking} is a reasoning enhancement technique built on LLM inference. MAD enables multiple agents to interact over several rounds. The goal is to encourage consensus on the same answer to a given query. A final decision is then produced through answer aggregation methods such as majority voting \citep{du2024improving}. In terms of output safety, MAD and RCS share a similar motivation. Both approaches introduce multiple models to improve the safety of the final output. Some MAD methods also adopt confidence scores to evaluate candidate answers proposed by other models \citep{chen-etal-2024-reconcile}. This mechanism is conceptually related to RCS. Compared to MAD, RCS follows a more fundamental strategy. RCS directly operates on underlying probability distributions. In contrast, MAD requires models to possess an explicit understanding of safety. This requirement limits robustness against unforeseen attacks. Moreover, RCS does not require interaction among models. This design avoids the conformity \citep{weng2025do} inherent in LLMs. Operating on underlying probability distributions also mitigates the impact of hallucination \citep{ji-etal-2023-towards} during reasoning. However, this design limits practical applicability. In many scenarios, models are accessed as black-box services, and probability distributions are not available. We argue that an intersection between MAD and RCS can improve the practical applicability of RCS. We continue to avoid direct interaction among models. Instead, each model outputs a score for each candidate output $y$ at every round. This score represents the level of support for $y$. All other components remain consistent with RCS. We refer to this variant as Practical Reliable Consensus Sampling (PRCS). More broadly, PRCS can be applied beyond safety. It can also enhance model reasoning performance in domains such as mathematics, medicine, and programming \citep{luo2025large}.

\section{Additional Experimental Results}
\label{app:results}
\begin{figure*}[h]
    \centering
    \begin{subfigure}{0.31\textwidth}
        \centering
        \includegraphics[width=\linewidth]{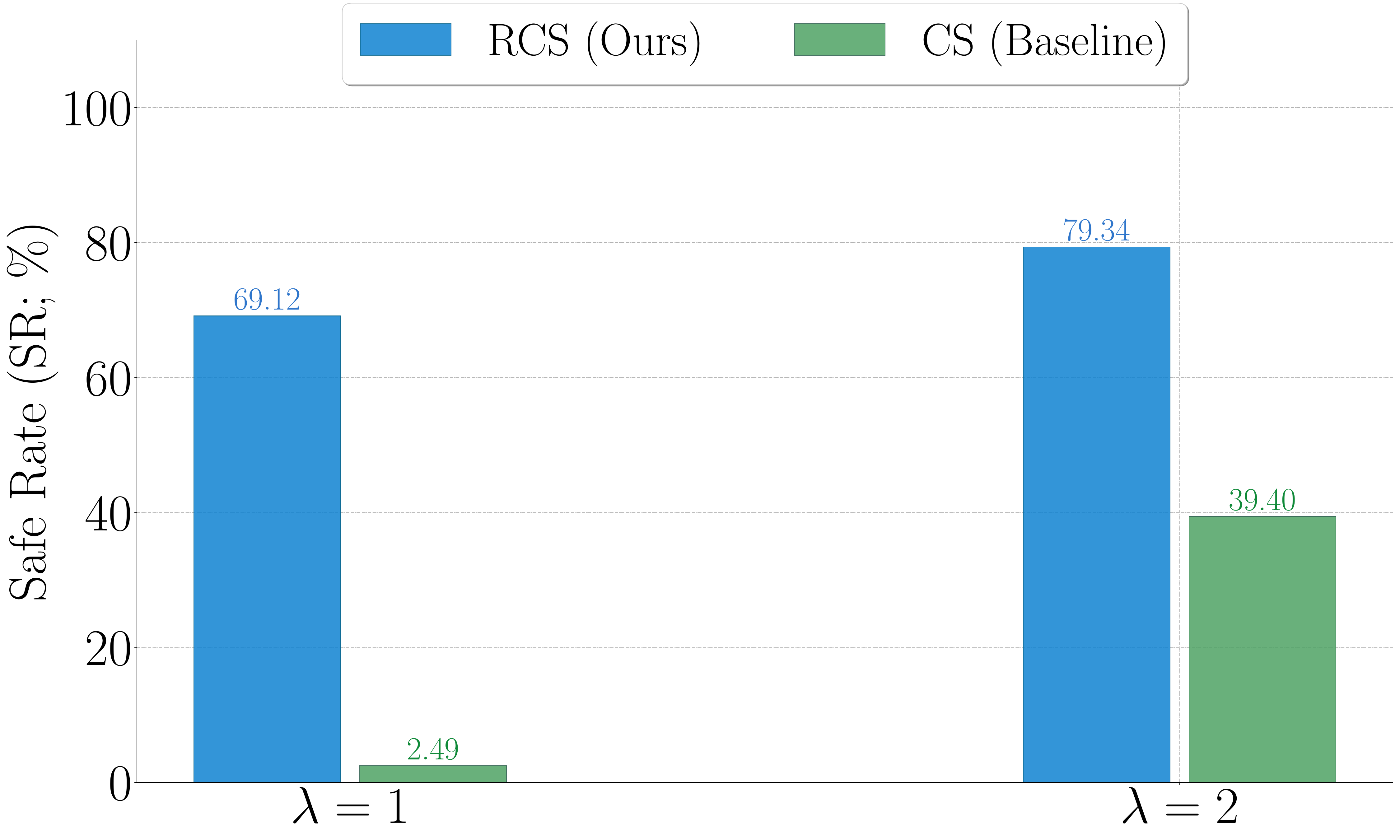}
        \caption{$n=5$.}
    \end{subfigure}
    \begin{subfigure}{0.31\textwidth}
        \centering
        \includegraphics[width=\linewidth]{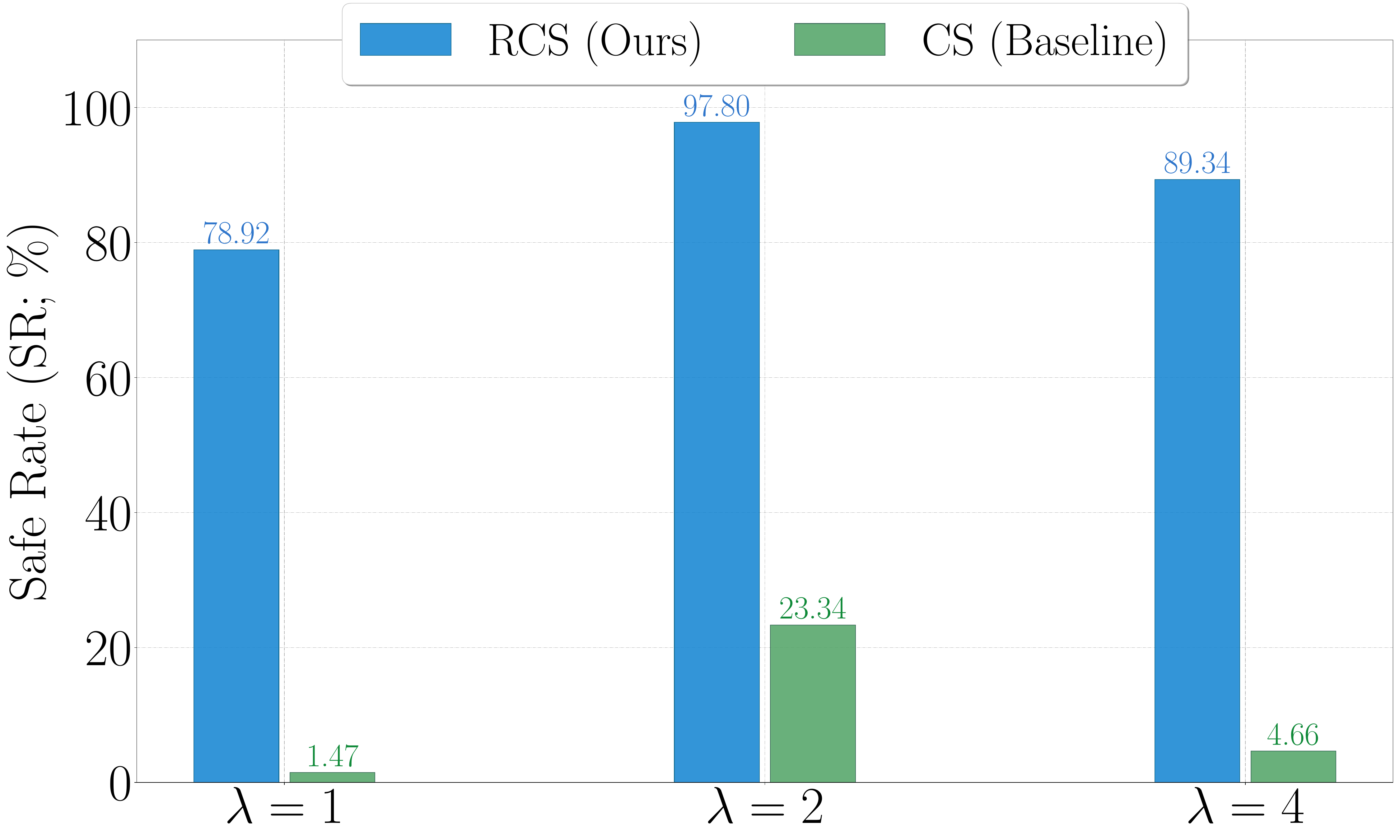}
        \caption{$n=9$.}
    \end{subfigure}
    \begin{subfigure}{0.31\textwidth}
        \centering
        \includegraphics[width=\linewidth]{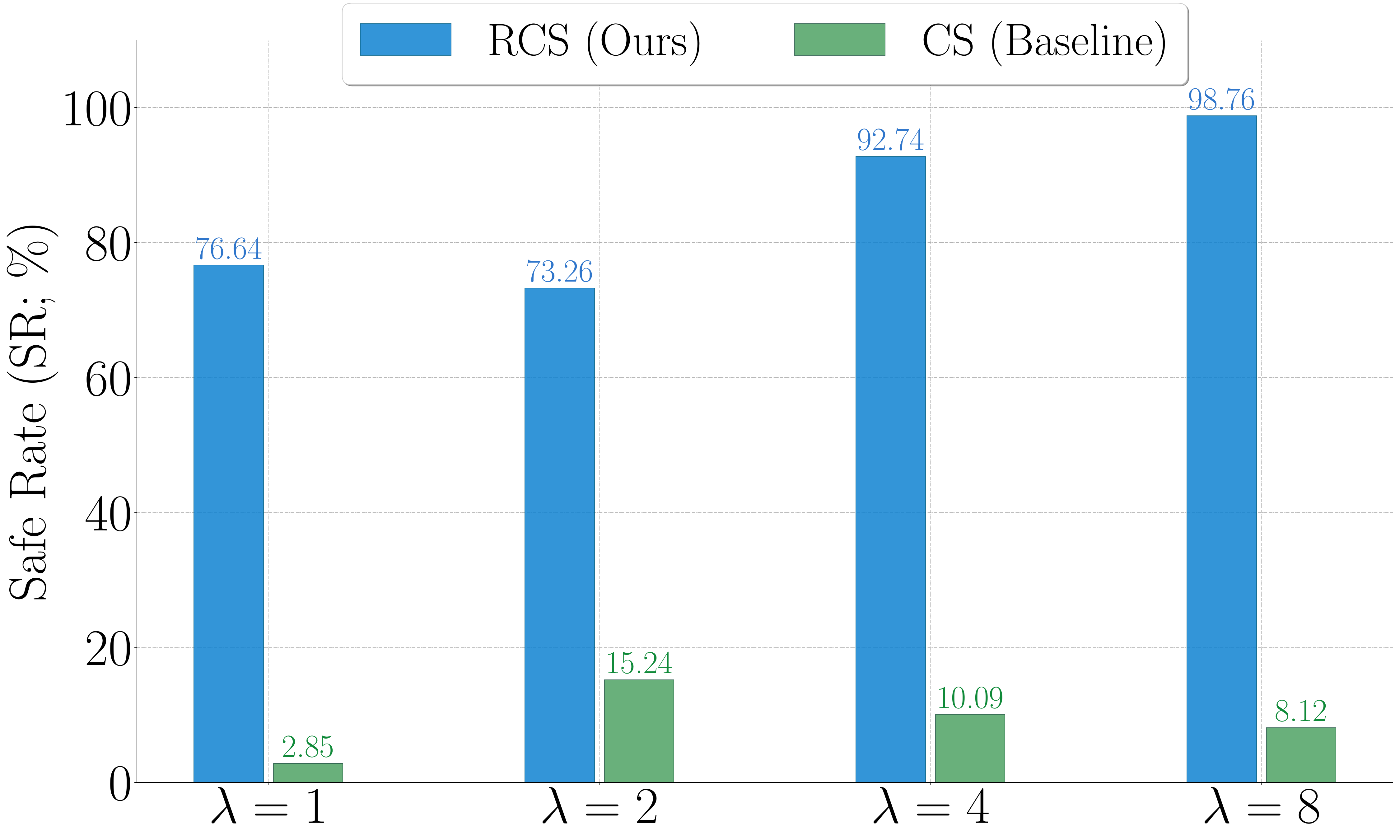}
        \caption{$n=17$.}
    \end{subfigure}
    \caption{Evaluation results for the safe rate when $f < \lceil \frac{n}{2} \rceil$.}
    \label{fig:sr_app}
    \vspace{-15pt}
\end{figure*}

\begin{figure*}[h]
    \centering
    \begin{subfigure}{0.31\textwidth}
        \centering
        \includegraphics[width=\linewidth]{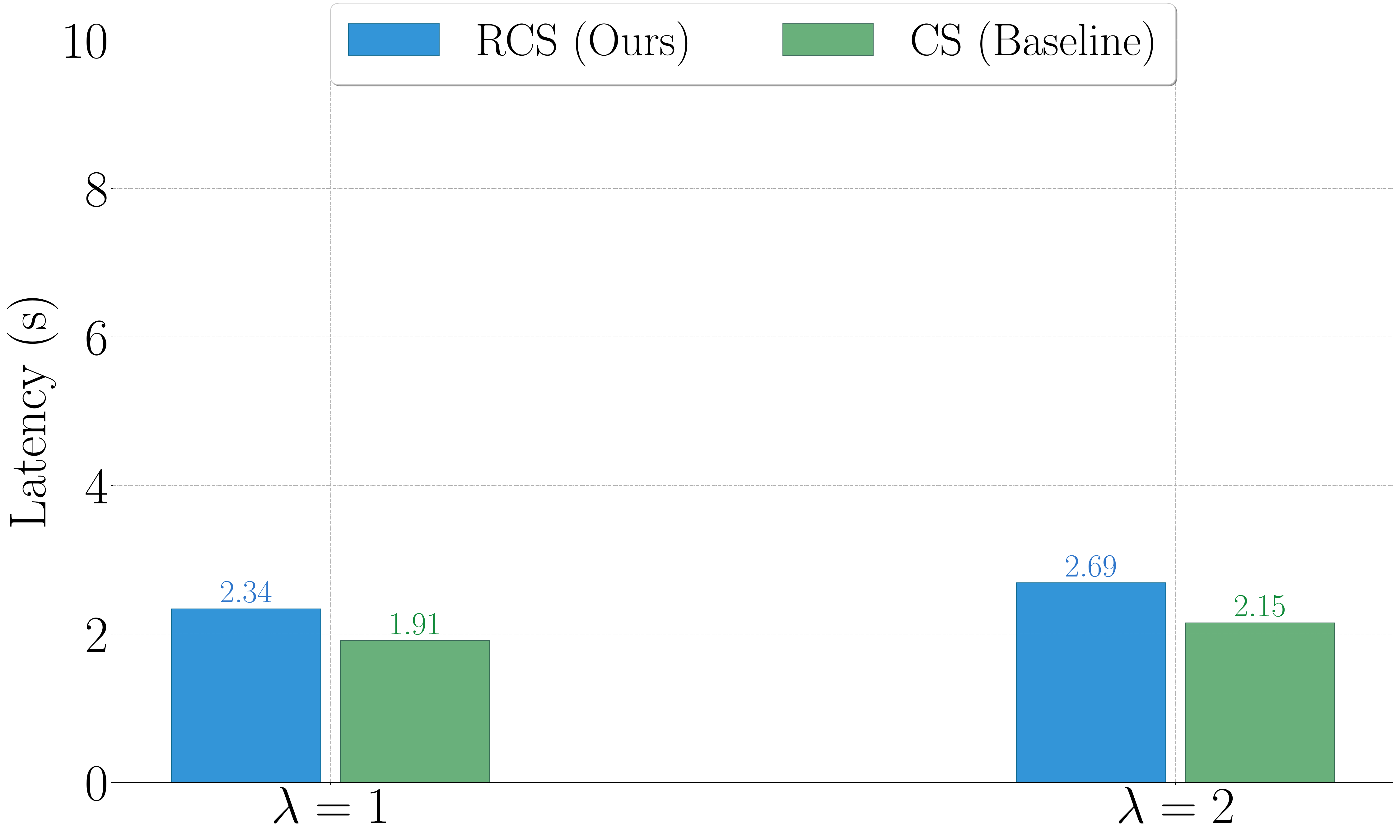}
        \caption{$n=5$.}
    \end{subfigure}
    \begin{subfigure}{0.31\textwidth}
        \centering
        \includegraphics[width=\linewidth]{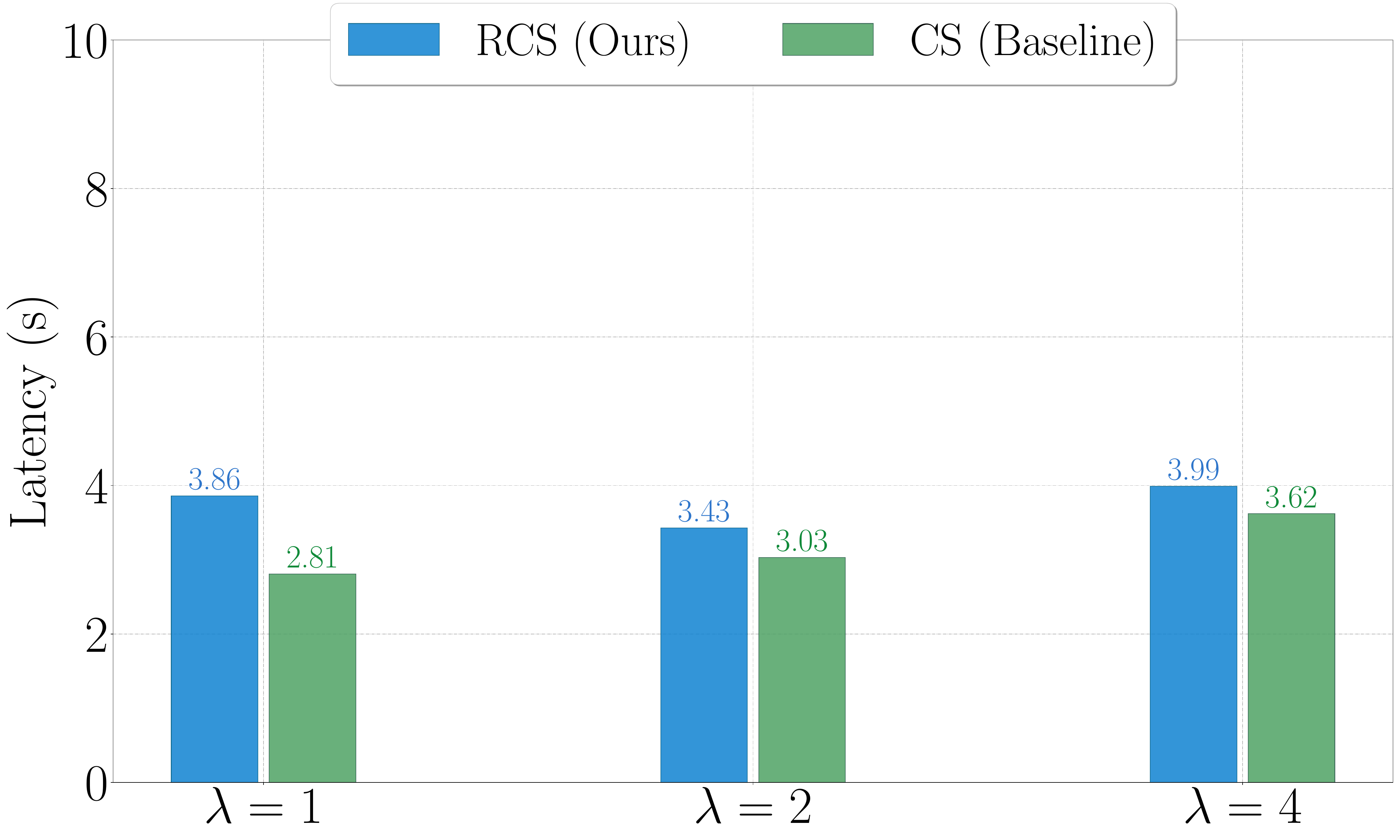}
        \caption{$n=9$.}
    \end{subfigure}
    \begin{subfigure}{0.31\textwidth}
        \centering
        \includegraphics[width=\linewidth]{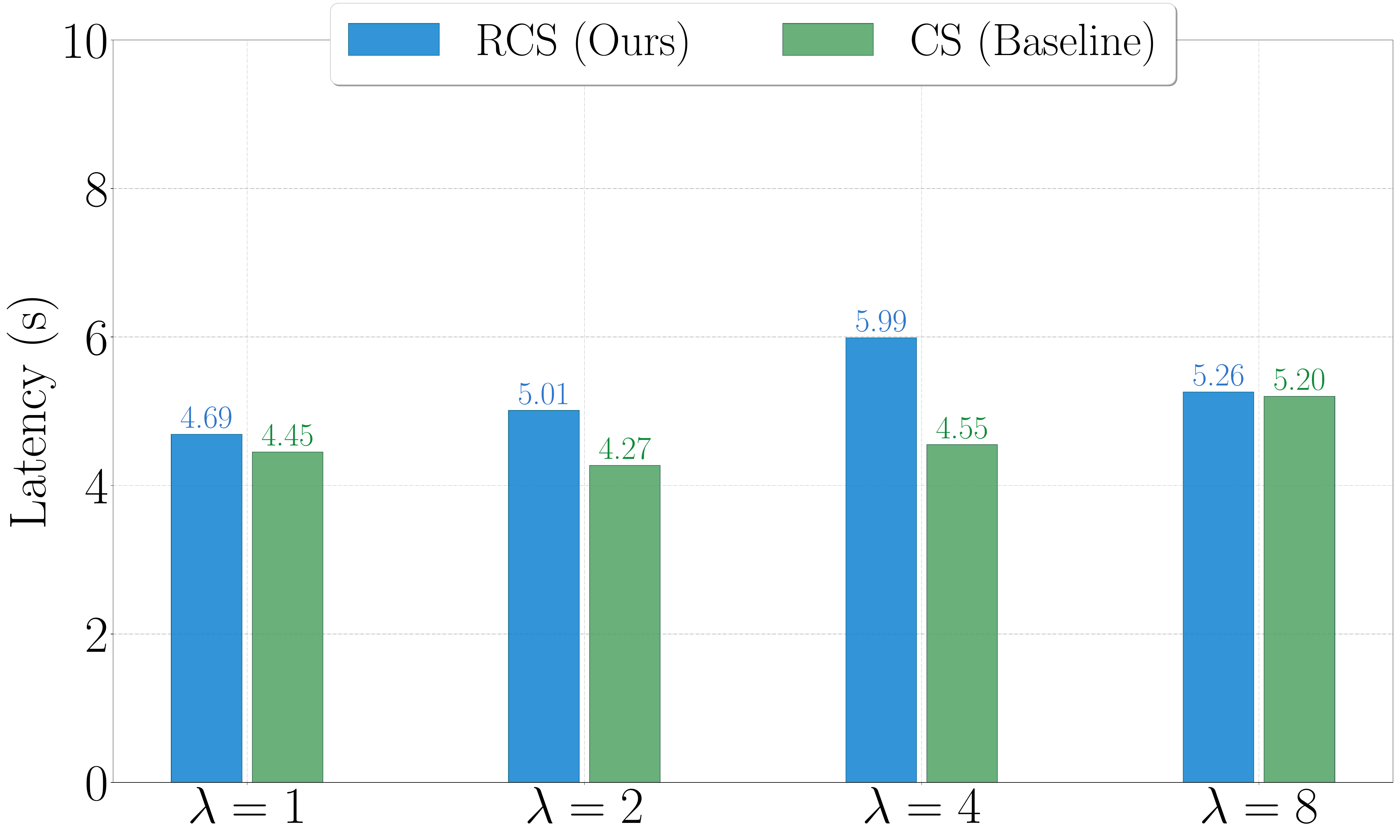}
        \caption{$n=17$.}
    \end{subfigure}
    \caption{Evaluation results for the latency when $f < \lceil \frac{n}{2} \rceil$.}
    \label{fig:TC_g_app}
\end{figure*}

\begin{figure*}[h]
    \centering
    \begin{subfigure}{0.31\textwidth}
        \centering
        \includegraphics[width=\linewidth]{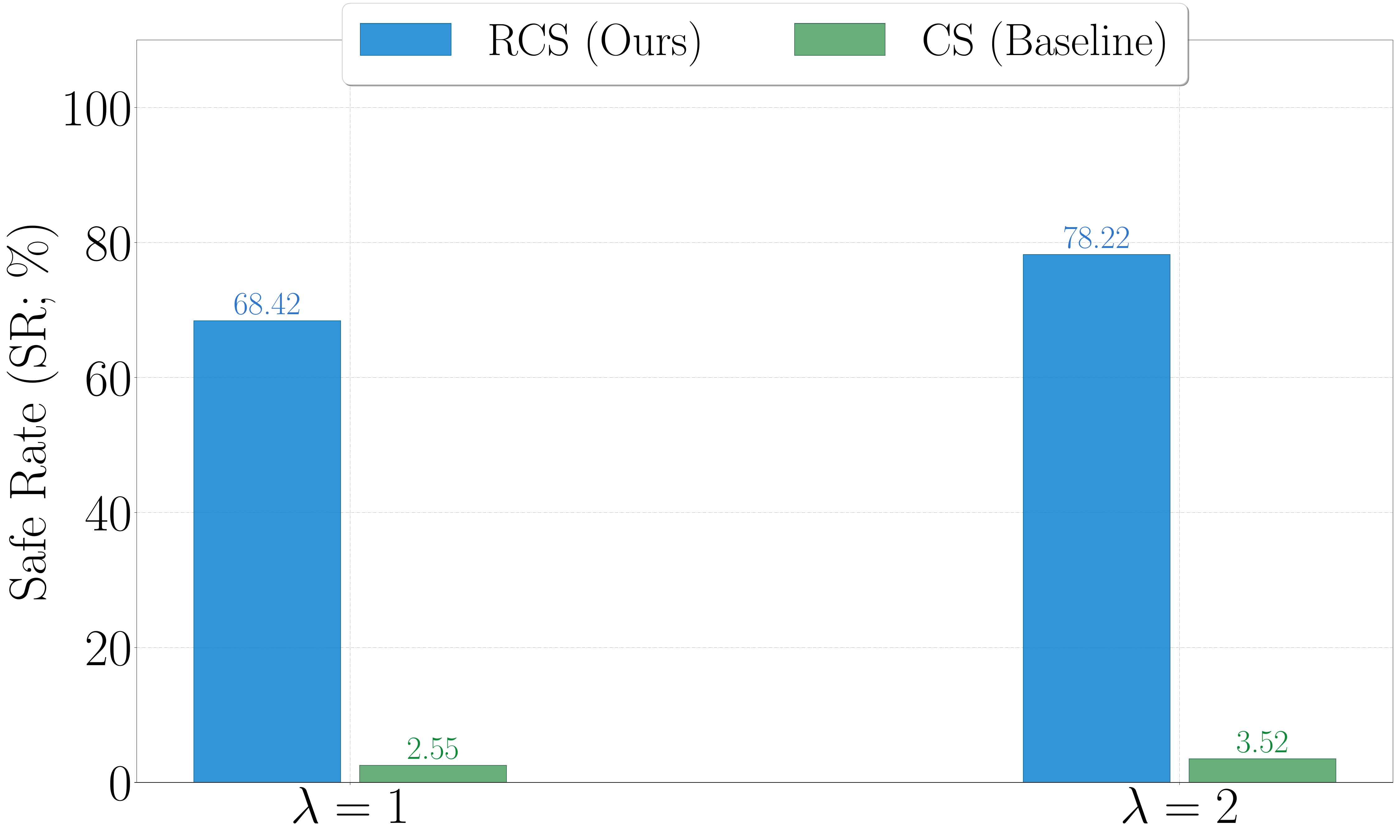}
        \caption{$n=5$.}
    \end{subfigure}
    \begin{subfigure}{0.31\textwidth}
        \centering
        \includegraphics[width=\linewidth]{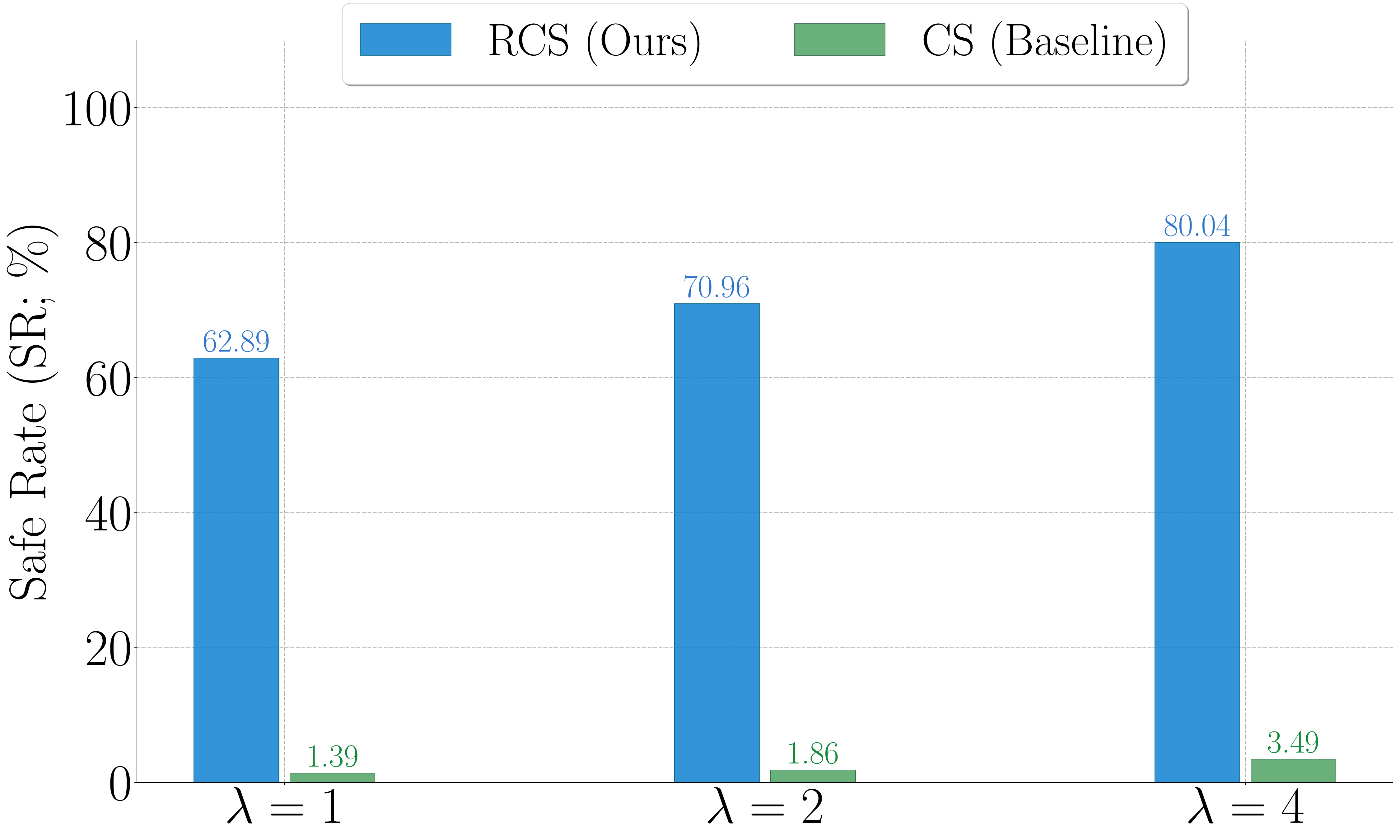}
        \caption{$n=9$.}
    \end{subfigure}
    \begin{subfigure}{0.31\textwidth}
        \centering
        \includegraphics[width=\linewidth]{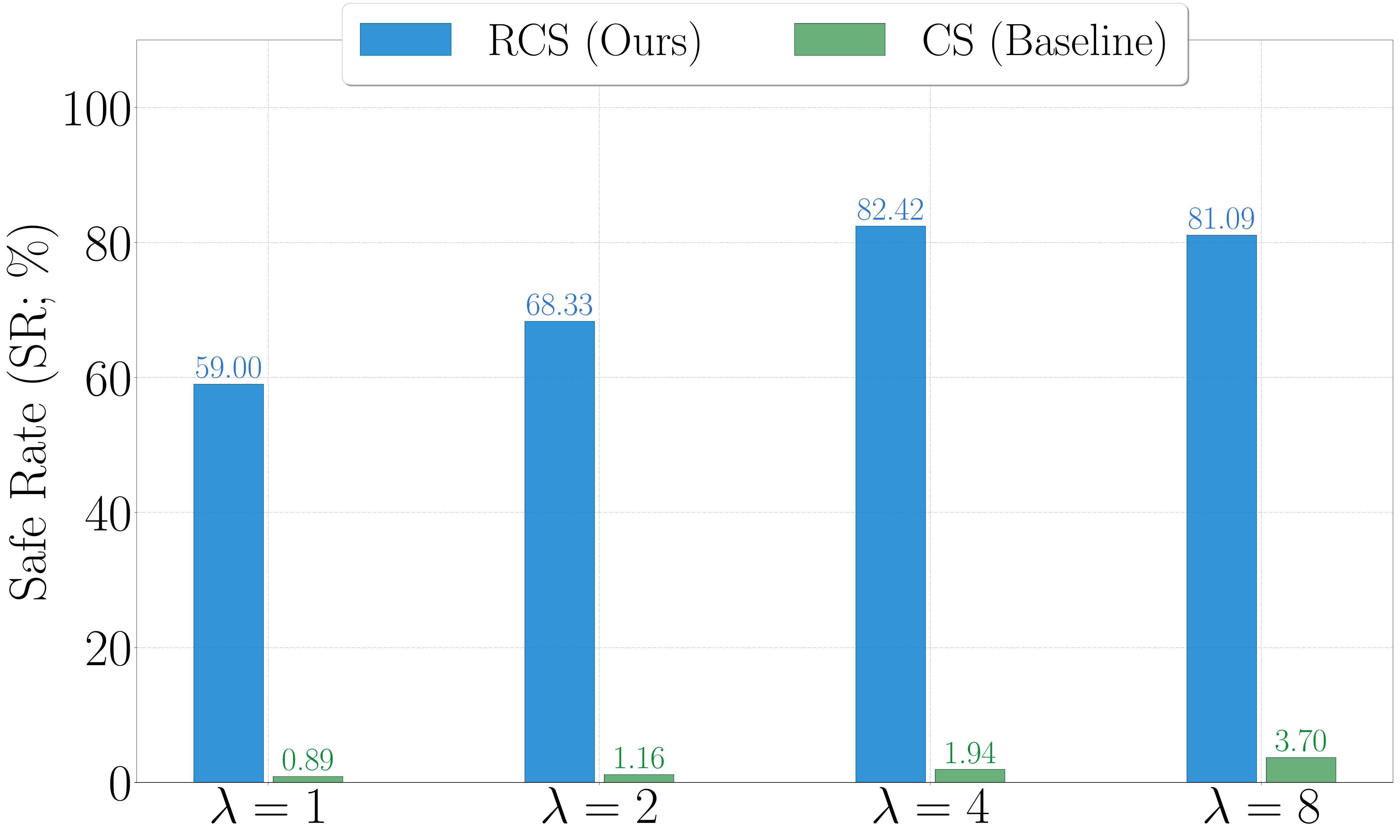}
        \caption{$n=17$.}
    \end{subfigure}
    \caption{Evaluation results for the safe rate when Byzantine models collude.}
    \label{fig:sr_m_app}
    \vspace{-15pt}
\end{figure*}

\newpage
\begin{figure*}[!]
    \centering
    \begin{subfigure}{0.32\textwidth}
        \centering
        \includegraphics[width=\linewidth]{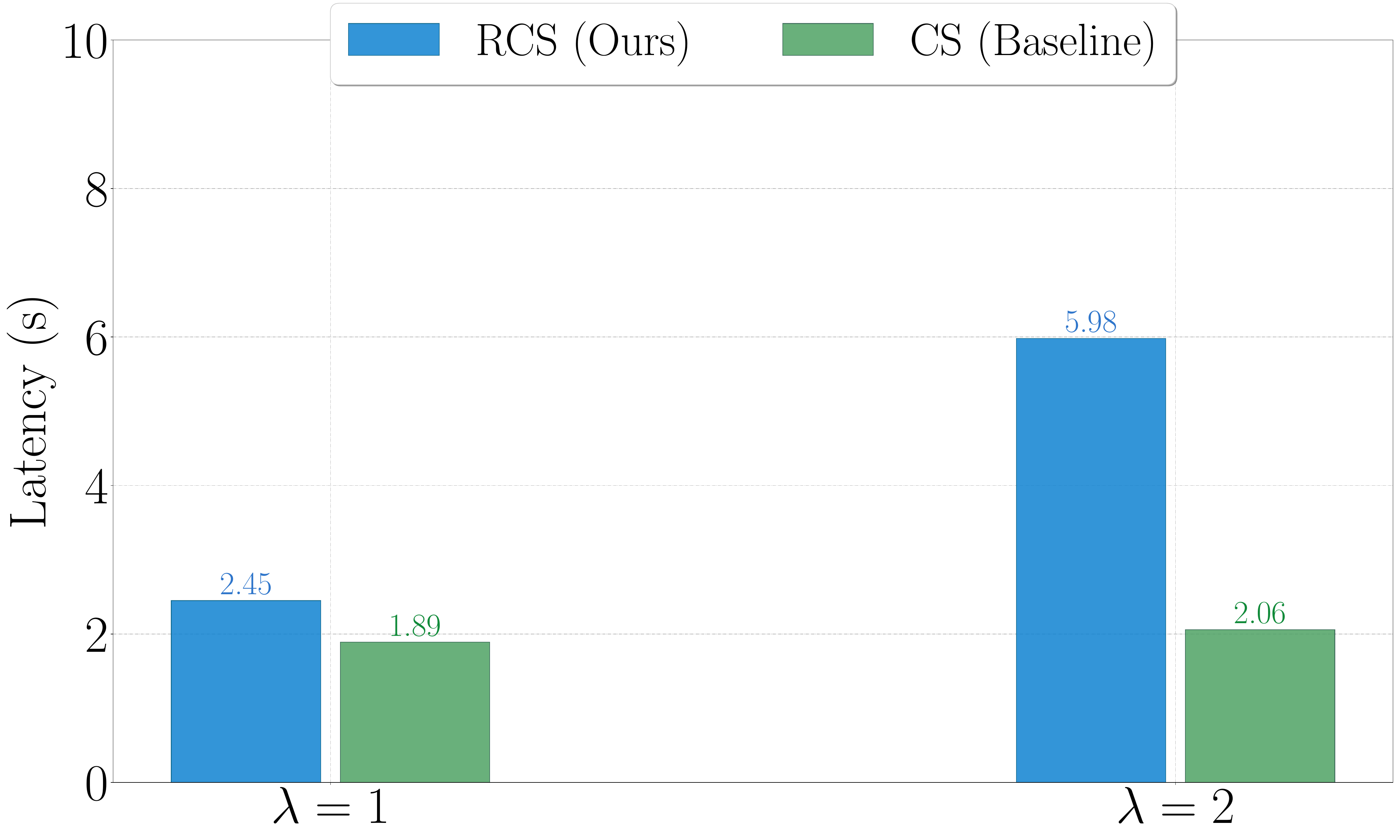}
        \caption{$n=5$.}
    \end{subfigure}
    \begin{subfigure}{0.32\textwidth}
        \centering
        \includegraphics[width=\linewidth]{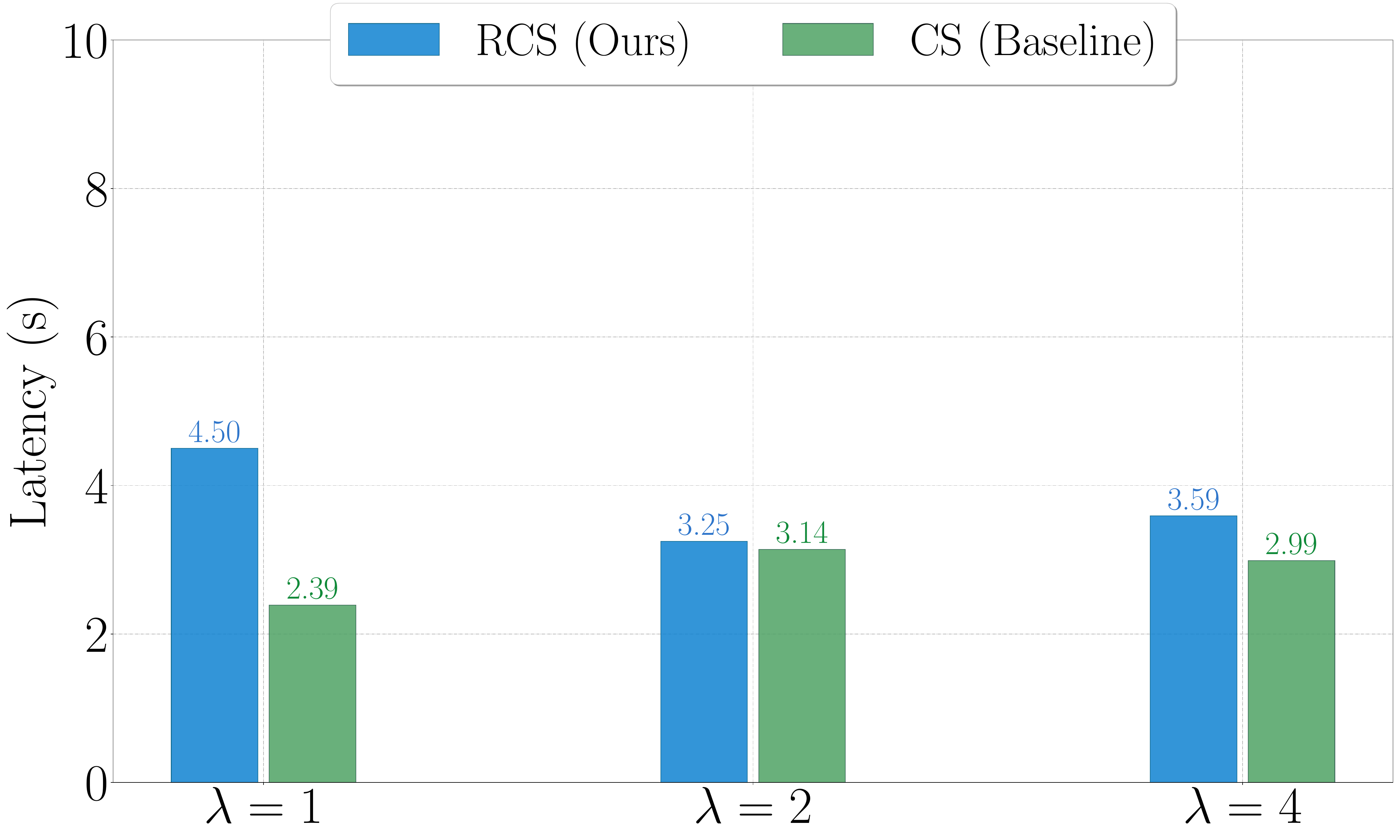}
        \caption{$n=9$.}
    \end{subfigure}
    \begin{subfigure}{0.32\textwidth}
        \centering
        \includegraphics[width=\linewidth]{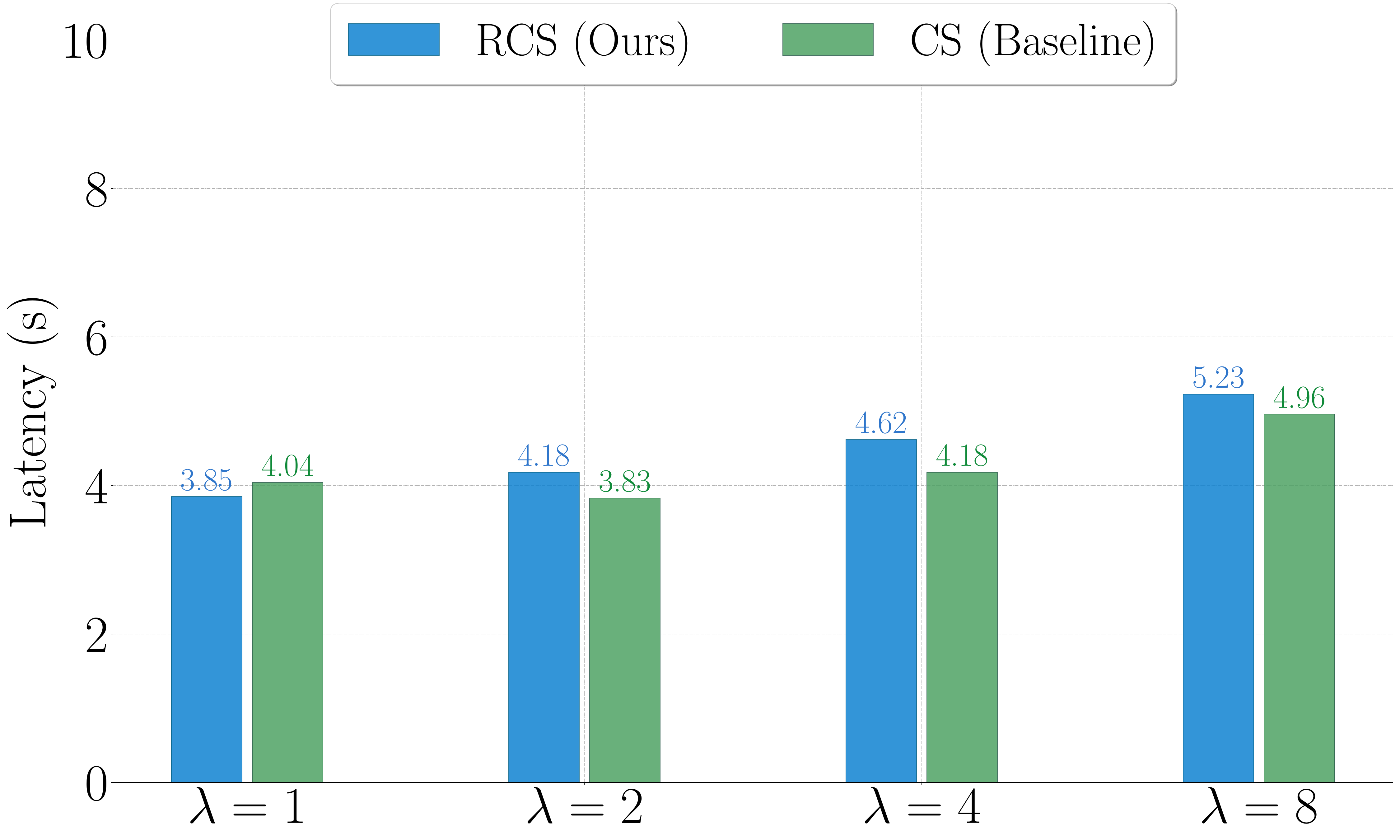}
        \caption{$n=17$.}
    \end{subfigure}
    \caption{Evaluation results for the latency when Byzantine models collude.}
    \label{fig:TC_m_app}
\end{figure*}

\begin{figure*}[!]
    \centering
    \begin{subfigure}{0.31\textwidth}
        \centering
        \includegraphics[width=\linewidth]{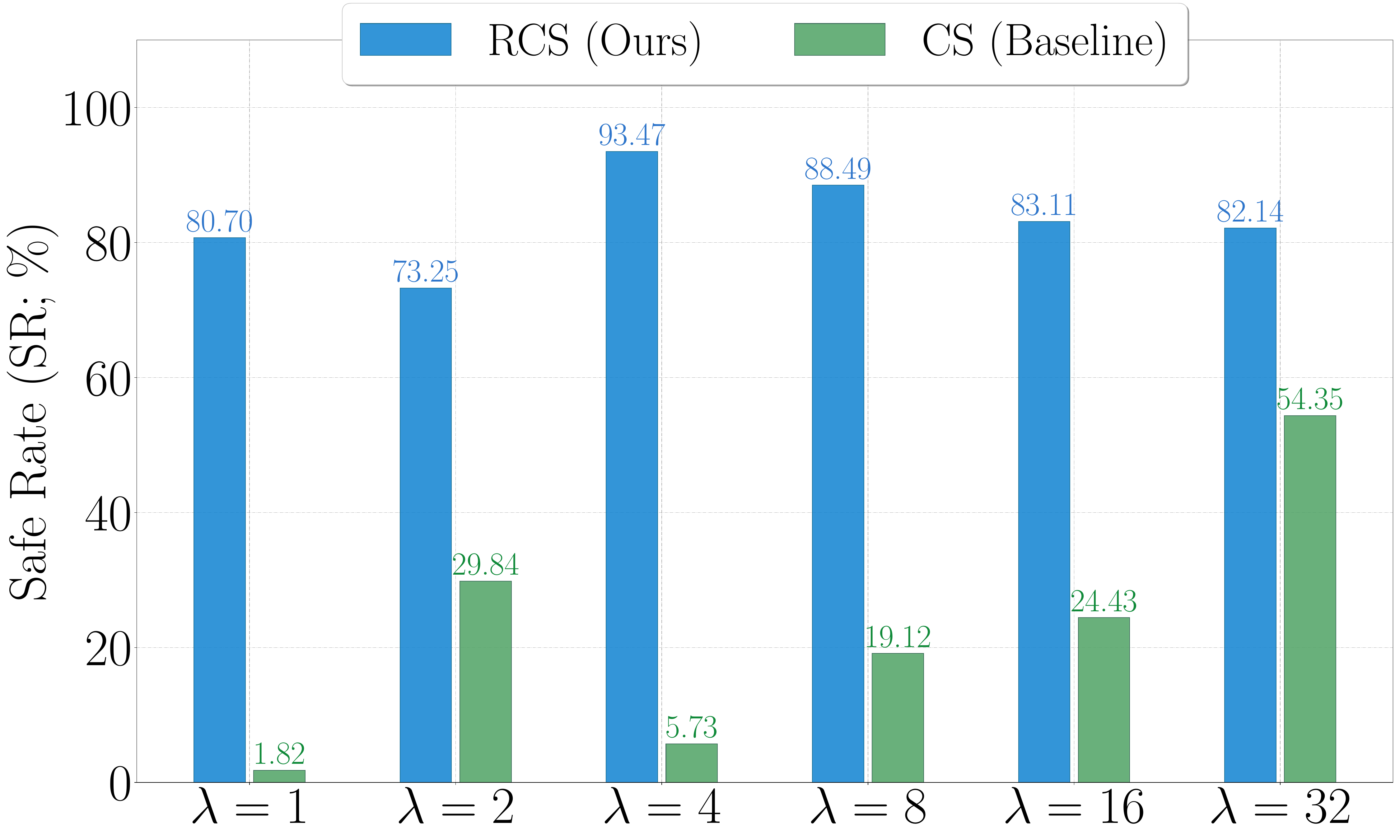}
        \caption{$n=34$.}
    \end{subfigure}
    \begin{subfigure}{0.31\textwidth}
        \centering
        \includegraphics[width=\linewidth]{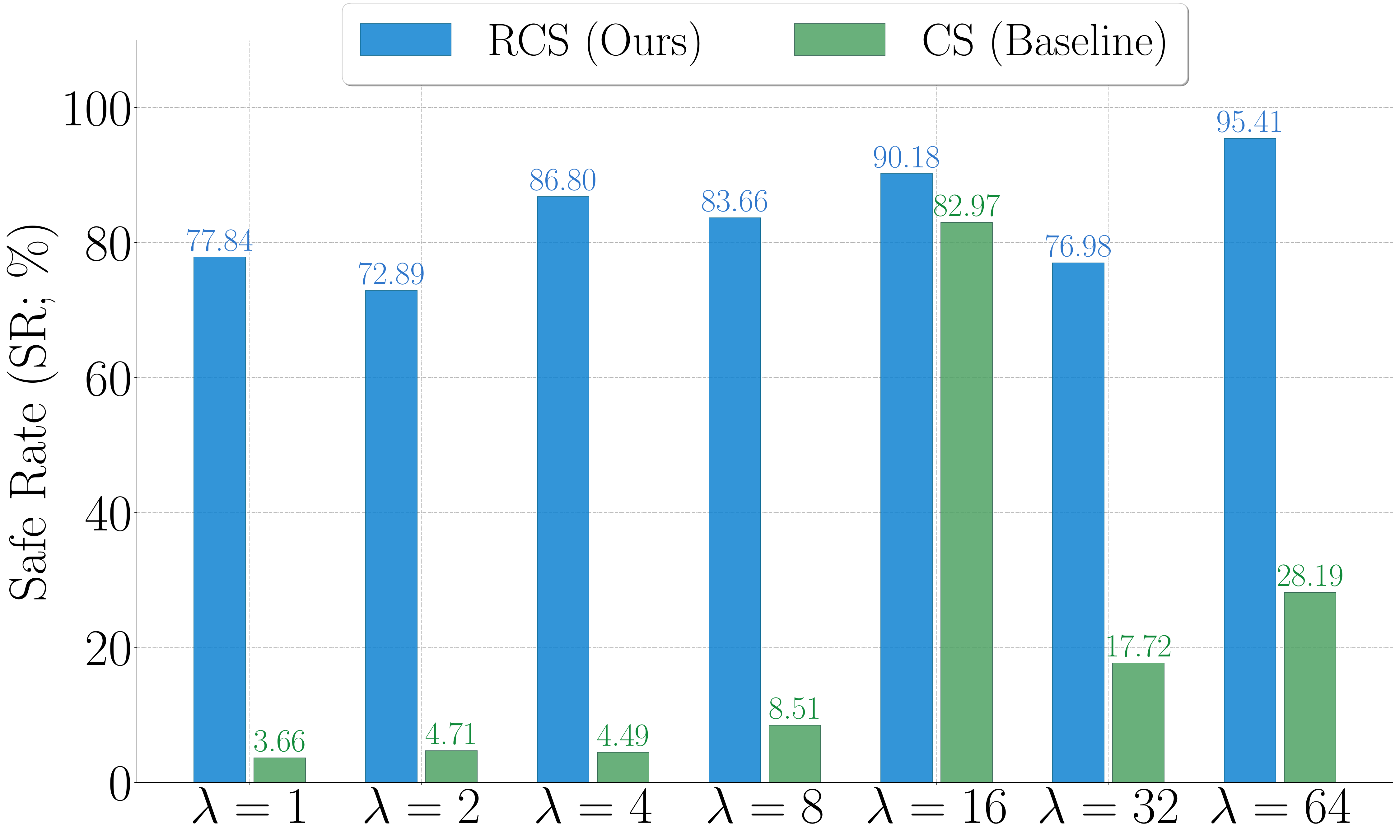}
        \caption{$n=66$.}
    \end{subfigure}
    \begin{subfigure}{0.35\textwidth}
        \centering
        \includegraphics[width=\linewidth]{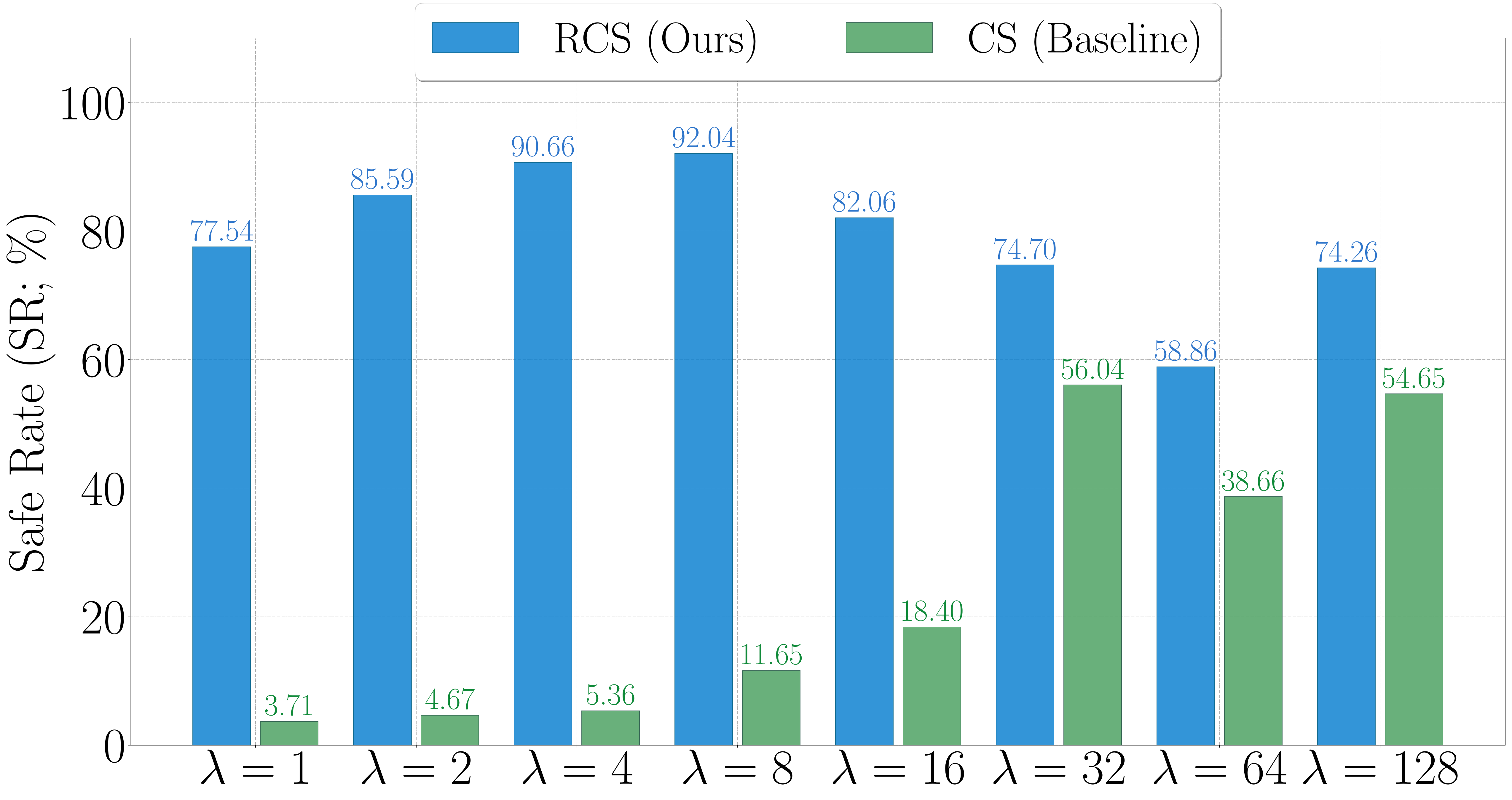}
        \caption{$n=130$.}
    \end{subfigure}
    \caption{Evaluation results for the safe rate when $f = \lceil \frac{n}{2} \rceil$.}
    \label{fig:SR_d_app}
\end{figure*}

\begin{figure*}[!]
    \centering
    \begin{subfigure}{0.31\textwidth}
        \centering
        \includegraphics[width=\linewidth]{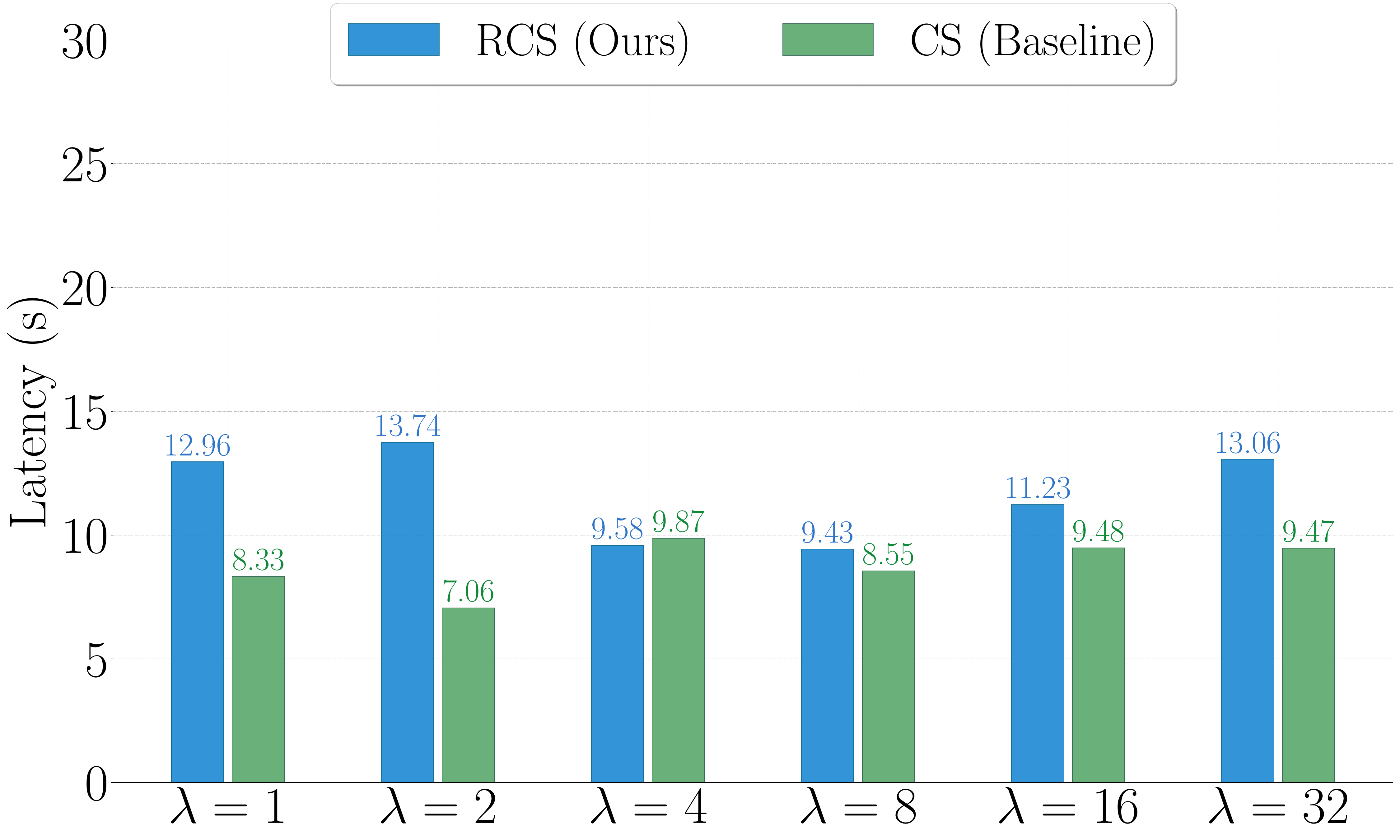}
        \caption{$n=34$.}
    \end{subfigure}
    \begin{subfigure}{0.31\textwidth}
        \centering
        \includegraphics[width=\linewidth]{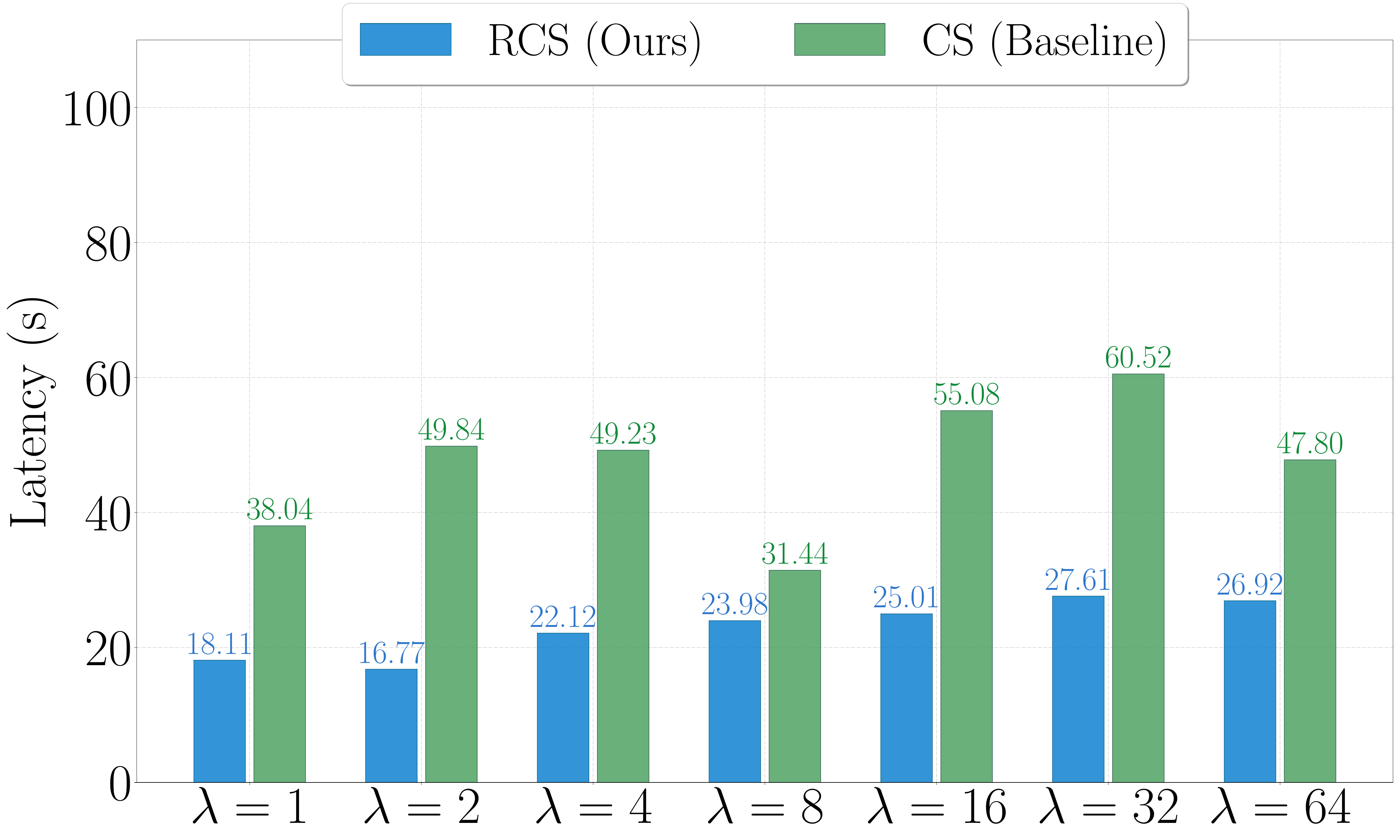}
        \caption{$n=66$.}
    \end{subfigure}
    \begin{subfigure}{0.35\textwidth}
        \centering
        \includegraphics[width=\linewidth]{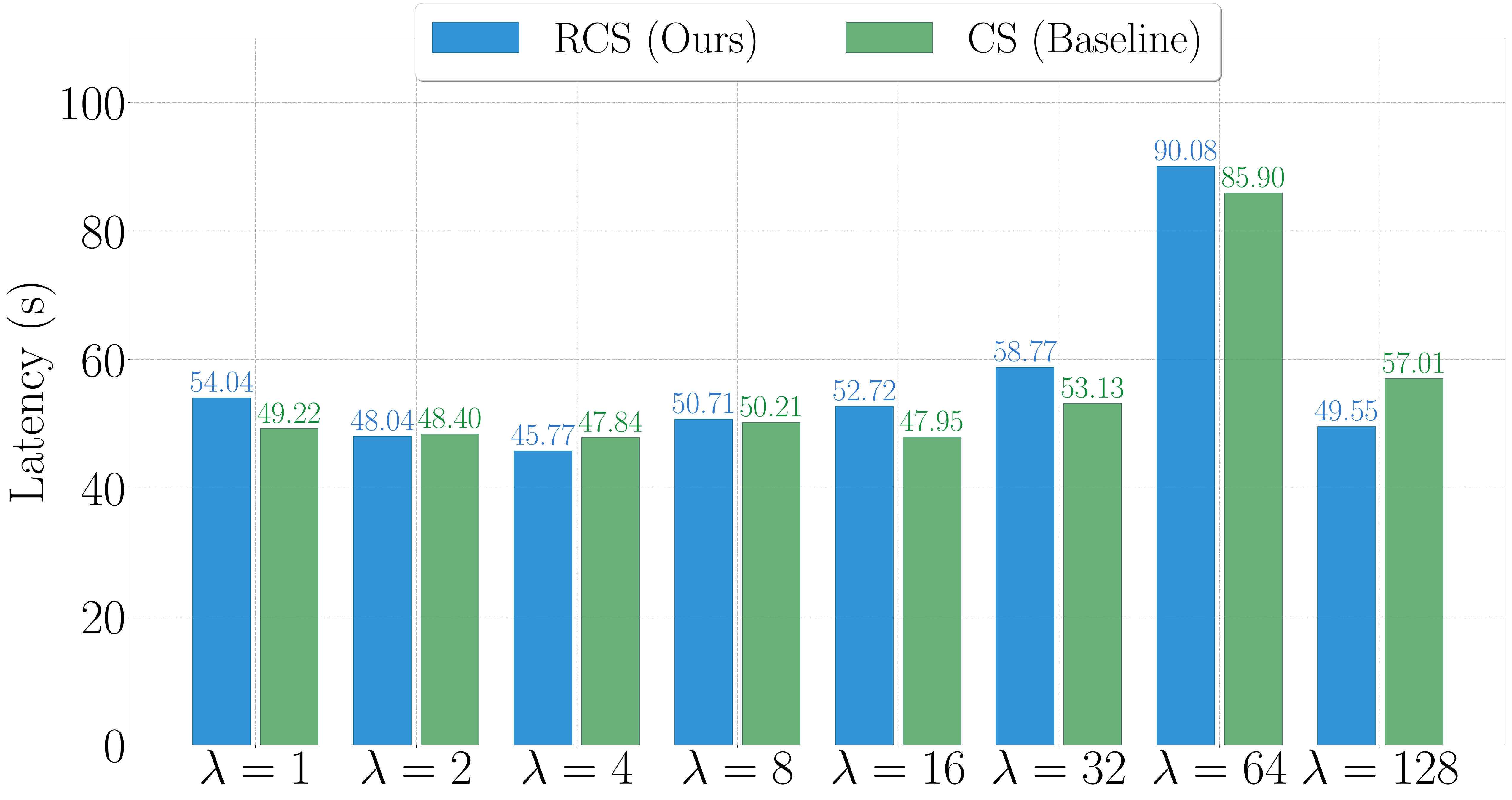}
        \caption{$n=130$.}
    \end{subfigure}
    \caption{Evaluation results for the latency when $f = \lceil \frac{n}{2} \rceil$.}
    \label{fig:TC_d_app}
\end{figure*}

\begin{figure}[!]
    \centering
\includegraphics[width=0.56\linewidth]{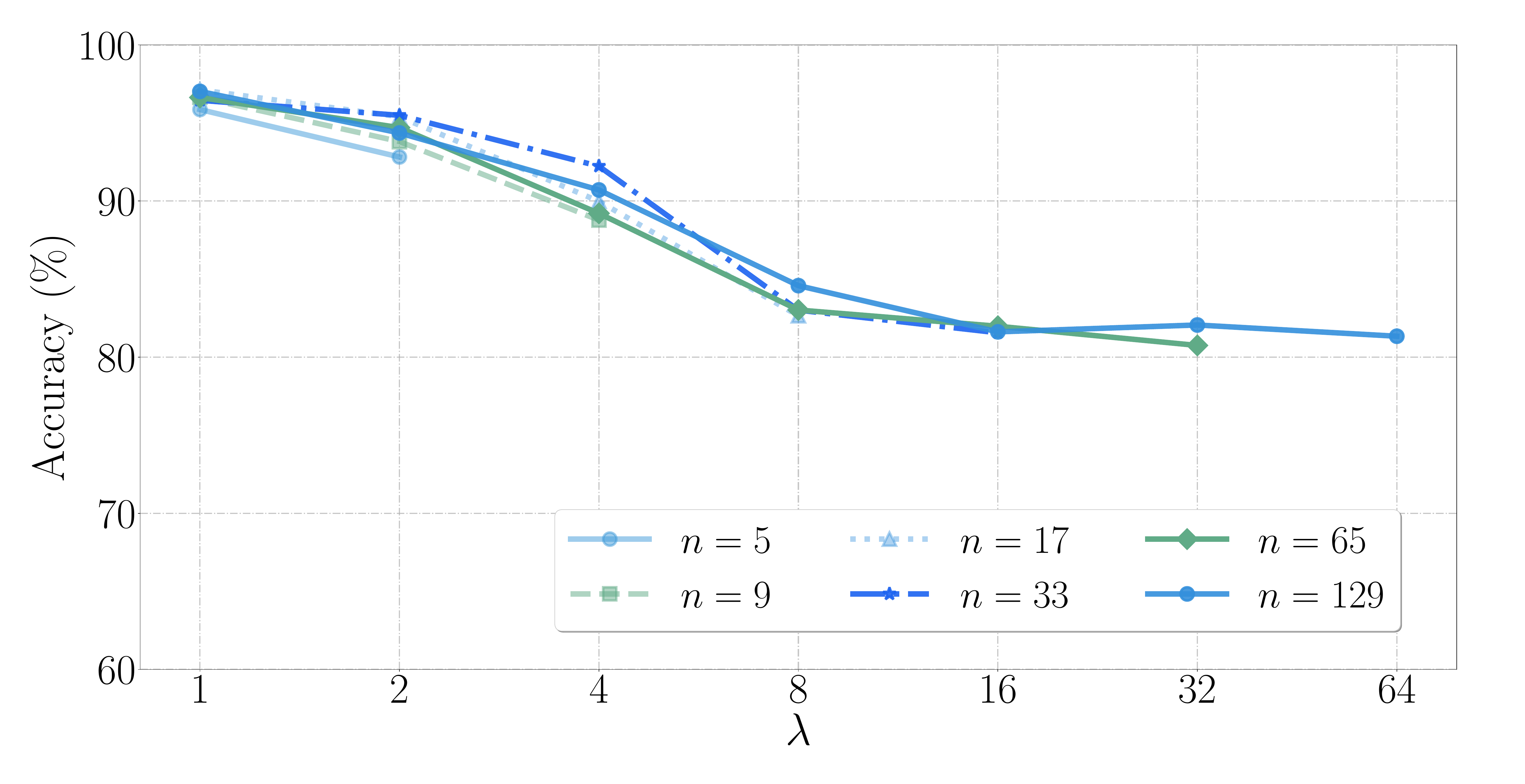}
    \caption{Accuracy of the feedback algorithm for different $\lambda$ and $n$ values when Byzantine models collude.}
    \label{fig:dr_m}   
    \vspace{-15pt}
\end{figure}

\section{Necessity Analysis of Liveness}
We further analyze the necessity of the liveness property. In real-world deployments, the goal of $\mathcal{MG}$ is not only to guarantee safety, but also to accomplish normal inference tasks such as mathematics, medical diagnosis, and programming. The objective is to maintain both functionality and safety, rather than prioritize safety as the sole consideration. As a result, in practical applications, $\mathcal{MG}$ must be able to perform valid reasoning on a wide variety of prompts. From the perspective of safety, abstention can be used to avoid unsafe responses. However, for regular inference tasks, abstention is not acceptable, as the response $\bot$ cannot adaptively provide reasonable outputs for unknown prompts. Therefore, liveness is necessary for $\mathcal{MG}$ from a practical application perspective. 

\end{document}